\documentclass[%
 aip,
 amsmath,amssymb,
 reprint,%
]{revtex4-1}

\usepackage{graphicx}
\usepackage{dcolumn}
\usepackage{bm}
\usepackage{lipsum}

\usepackage[utf8]{inputenc}
\usepackage[T1]{fontenc}
\usepackage{mathptmx}
\usepackage{xcolor}
\usepackage{mathrsfs}

\begin{document}

\preprint{AIP/123-QED}

\title{Elastic instabilities and bifurcations in flows of wormlike micellar solutions past  single and two vertically aligned microcylinders: Effect of blockage and gap ratios}
\author{Mohd Bilal Khan}
\author{C. Sasmal}%
 \email{csasmal@iitrpr.ac.in}
\affiliation{ 
Soft Matter Engineering and Microfluidics Lab, Department of Chemical Engineering, Indian Institute of Technology Ropar, Punjab, India-140001.
}



\date{\today}

\begin{abstract}
This study presents an extensive numerical investigation on the flow characteristics of wormlike micellar solutions past a single and vertically aligned two microcylinders placed in a microchannel in the creeping flow regime. The rheological behaviour of the micellar solution is realized based on the two-species Vasquez-Cook-McKinley (VCM) constitutive model, which takes into account of both the breakage and reformation dynamics of micelles. For the case of single microcylinder, as the blockage ratio (ratio of the cylinder diameter to that of the channel height) is gradually varied, we find the existence of a flow bifurcation in the system, and also a gradual transition for a range of flow states, for instance, steady and symmetric or Newtonian like, steady and asymmetric, unsteady periodic and asymmetric, unsteady quasi-periodic and asymmetric, and finally, unsteady quasi-periodic and symmetric. For the case of two microcylinders, we observe the presence of three distinct flow states in the system, namely, diverging (D), asymmetric-diverging (AD) and converging (C) states as the intercylinder spacing in between the two cylinders is varied. Similar types of flow states are also observed in the recent experiments dealing with wormlike micellar solutions. However, we show that either this transition from one flow state to another in the case of a single microcylinder or the occurrence of any flow state in the case of two microcylinders, is strongly dependent upon the values of the Weissenberg number and the non-linear VCM model parameter $\xi$, which basically indicates how easy or hard to break a micelle. Based on the results and discussion presented herein for the single and two microcylinders, we ultimately provide the explanation for the formation of preferential paths or lanes during the flow of viscoelastic fluids through a porous media, which was seen in many prior experiments in the creeping flow regime.  
\end{abstract}

\maketitle

\section{\label{Intro}Introduction}
Addition of a small amount of highly flexible surfactant molecules into a solvent like water greatly influences the flow characteristics of the resulting solution in a broad-spectrum of measurable scales.  Beyond a critical concentration, these amphiphilic surfactant molecules spontaneously self-assemble and form a large aggregate called micelles, which can be of different shapes like spherical, ellipsoidal, wormlike, or lamellae~\cite{dreiss2007wormlike,dreiss2017wormlike}. Further increasing the surfactant concentration leads to the entanglement of these micelles, thereby originating complex viscoelastic properties~\cite{yang2002viscoelastic,walker2001rheology}. However, the rheological behaviour of these micellar solutions, particularly wormlike micellar solutions, is found to more complex than that seen for polymer solutions or melts under otherwise identical conditions~\cite{rothstein2008strong,rothstein2003transient,berret1997transient}. This is because of the fact that these wormlike micelles can undergo continuous scission and reformation in a flow field, which is unlikely to happen for polymers due to the presence of a strong covalent backbone. Due to the presence of interesting rheological properties, these micellar solutions are widely used in many industrial applications, such as in the petroleum industry in the enhanced oil recovery process, as drag reducing agent, in cosmetics and pharmaceutical industries, in coating and paints industries, in biomedical applications, etc~\cite{schramm2000surfactants,mobius2001surfactants,raffa2015polymeric}. Therefore, a detailed understanding of the complex flow behaviour of these micellar solutions is very much needed for their better applications.  

One of the examples wherein the complex flow behaviour of micellar solutions can be seen is the flow through a porous media. In many experiments, it has been found that the micellar solution selects a preferential path or lane during the flow through a porous media. For instance, De et al.~\cite{de2018flow} observed the formation of lanes when a micellar solution comprising of cetyl tri‐methyl ammonium bromide (CTAB) and sodium salicylate (NaSal) flows through a model porous media consisting of a microchannel with cylindrical pillars placed in it. In another study~\cite{de2017lane}, they found a similar formation of lanes and their path switching phenomena when dealing with a hydrolyzed polyacrylamide (HPAM) polymer solution. Muller et al.~\cite{muller1998optical} also noticed the same phenomena in polyalphaolefine polymer solution flowing in a model porous medium consisting of a glass pipe filled with Duran glass spheres. They further noted spatial and temporal variations of these preferential paths in the porous media.  Recently, both Walkama et al.~\cite{walkama2020disorder} and Eberhard et al.~\cite{eberhard2020mapping} also showed the formation of these lanes in both ordered and disordered model porous structures during the flow of a high molecular weight polyacrylamide (PAA) and xanthan gum polymer solutions, respectively.        

To understand such complex flow behaviour of either micellar or polymer solutions in a porous media, it is always better to start with a simple system consisting of a single microcylinder placed in a microchannel. This simple benchmark system creates a non-homogeneous flow field in the system, which in turn, facilitates the understanding of the flow behaviour of various complex fluids. This ultimately leads to a better understanding of the flow behaviour in a more complex system. For this reasoning, a significant amount of studies, comprising of both experiments and numerical simulations, have been carried out on this benchmark system both for polymer~\cite{alves2001flow,mckinley1993wake,hu1990numerical,shiang1997viscoelastic,qin2019upstream} as well as micellar~\cite{moss2010flow,zhao2016flow,haward2019flow,khan2020effect} solutions. Some interesting flow physics have been found from these studies which were not seen in simple Newtonian fluids under otherwise identical conditions. For instance, the emergence of an elastic instability~\cite{qin2019upstream} and flow bifurcation~\cite{haward2019flow} have been found in this model geometry.  

Although the geometrical configuration of this model system is simple, the flow dynamics within it can be greatly altered either by changing the blockage ratio (ratio of the cylinder diameter to the channel height) or by placing another microcylinder next or above or bottom to the existing cylinder with various intercylinder spacings. For instance, both Moss and Rothstein~\cite{moss2010flow} and Zhao et al.~\cite{zhao2016flow} found that the onset of the elastic instability in CPyCl (cetylpyridinium chloride)/NaSal and CTAB/SHNC (3-hydroxy naphthalene-2-carboxylate) micellar solutions were delayed as the blockage ratio was decreased. Furthermore, Zhao et al.~\cite{zhao2016flow} observed a broad spectrum of flow states in this model geometry as the blockage ratio and Weissenberg number were varied, for instance, Newtonian like, bending streamlines, vortex growth upstream, unsteady downstream, chaotic upstream and three-dimensional time dependent. Recently, Varchanis et al.~\cite{varchanis2020asymmetric} conducted both experiments using polyethylene oxide (PEO) polymer solution and numerical simulations using the linear Phan-Thein-Tanner (I-PTT) constitutive model over a wide range of the blockage ratio. They found an existence of the supercritical and subcritical pitchfork bifurcations in the flow field as the blockage ratio was varied, and also observed no bifurcation in the flow for certain ranges of the blockage ratio. 

Apart from the influence of the blockage ratio, the placing of another microcylinder in the channel can also greatly modify the flow field in this model geometry. For example, Haward et al.~\cite{haward2018steady} experimentally found a significant modification in the flow field in between the two microcylinders than that seen for the single microcylinder case, particularly at high Weissenberg numbers. Varshney and Steinberg~\cite{varshney2017elastic} found an increase in the vortex growth in between the two microcylinders. This is in stark contrast to the findings of the suppression of a vortex by the polymer additives into a Newtonian solvent~\cite{cressman2001modification,zhu2019vortex}. Both these studies used a polymer solution in their experiments wherein two microcylinders were placed horizontally side-by-side. Recently, Hopkins et al.~\cite{hopkins2020tristability} performed experiments using CPyCl/NaSal micellar solution for the flow past two microcylinders placed vertically side-by-side over a broad range of the intercylinder gaps and Weissenberg numbers. This experimental study, performed for the first time for this geometry, found the existence of three stable flow states in the system depending upon the values of the intercylinder gap and Weissenberg number, namely,  diverging (D) state in which all of the fluid preferably passes through the gaps in between the channel walls and cylinder surface, asymmetric-diverging (AD) state in which the fluid prefers to pass through either the gap in between the upper channel wall and top cylinder surface or the lower channel wall and bottom cylinder surface, and converging (C) state in which most of the fluids pass through the gap in between the two cylinders. They presented a phase diagram on the existence of all these flow states as a function of the intercylinder gap and Weissenberg number, and also found a critical value of the intercylinder gap at which all these three states, namely, D, AD and C co-exist together, thereby showing the existence of a tristable state in viscoelastic fluids for the first time.

All these aforementioned studies demonstrate that the flow physics past a microcylinder confined in a channel can become increasingly complex if one changes either the blockage ratio or places an additional microcylinder in it. This is primarily due to the variation of the extent of shear and extensional flow fields in the domain, and due to the interaction of the elastic stresses generated around the microcylinders. However, it can be seen that most of these investigations are experimental, and in comparison to this, a very few numerical studies have been carried out~\cite{varchanis2020asymmetric}. Furthermore, these numerical simulations are based on the single-species viscoelastic constitutive equations, thus restricting their applicability to only polymer solutions in which breakage and reformation dynamics are absent unlike wormlike micellar solutions. Therefore, these widely used single-species viscoelastic constitutive equations sometimes unable to predict some typical flow physics happening in wormlike micellar solutions. For instance, many experimental studies have found an existence of unsteady motion of a sphere falling freely in wormlike micellar solutions in the creeping flow regime once the Weissenberg exceeds a critical value~\cite{mohammadigoushki2016sedimentation,chen2004flow}. It was predicted experimentally that this motion was due to the breakage of long and stretched micelles downstream of the sphere, resulting from an increase in the extensional flow strength. Only recently~\cite{sasmalJFM}, it has been proven that this motion is, indeed, due to the breakage of micelles downstream of the sphere using the two-species Vasquez-Cook-McKinley (VCM) model~\cite{vasquez2007network}. This model considers the wormlike micelles as an elastic segment composed of Hookean springs, which all together form an elastic network that can continuously break and reform in a flow field. The breaking and reforming processes of this model were incorporated based on the discrete and simplified version of Cate's reversible breaking theory for wormlike micelles~\cite{cates1987reptation}. According to this model, a long micelle of length $L$ is likely to break in the middle into two short micelles of equal length of $L/2$, and two short micelles can also recombine into a long micelle. This is opposed to the Cate's original theory in which a long micelle can break at any point along their length with equal probability and also micelles of any length can join together to form a long micelles. However, the simplification adopted for the breakage and reformation dynamics in the VCM model makes an easy implementation in any CFD platform to simulate the complex flows of micellar solutions, and it also allows to capture the temporal and spatial variations in the number density of short and long micelles. 

The VCM model efficiently captures all the typical flow characteristics of wormlike micellar solutions like shear thinning, shear banding, extensional hardening and subsequent thinning, etc. in homogeneous viscometric flows~\cite{pipe2010wormlike,zhou2014wormlike}. For non-viscometric flows, the VCM model also successfully predicts many experimental observations seen in flows through complex geometries, for instance, the formation of a lip vortex in a microfluidic cross-slot cell~\cite{kalb2017role,kalb2018elastic}, flow characteristics in a micropore with step expansion and contraction~\cite{sasmal2020flow}, transient evaluation of the velocity profiles in a Taylor-Couette flow~\cite{mohammadigoushki2019transient}, etc. Only recently, the flow characteristics of WLM solutions through the benchmark system of a microcylinder confined in a channel at a fixed blockage ratio have been studied based on this VCM model by us in our earlier study~\cite{khan2020effect}. In this investigation, likewise the experiments~\cite{moss2010flow,zhao2016flow}, we have also observed the emergence of an elastic instability in the system once the Weissenberg exceeds a critical value. Furthermore, we have shown that this instability is greatly influenced by the non-linear VCM model parameter $\xi$ which basically indicates how easy or hard to break a micelle. However, still, there is a gap of knowledge present in the literature, in particular, for the flow past two vertically aligned microcylinders which may facilitate the understanding of the formation of preferential paths or lanes during the flow of viscoelastic fluids in a porous media.

Therefore, the aim of this study is threefold: firstly, we aim to numerically investigate how the blockage ratio would tend to influence the flow dynamics of a micellar solution past a single microcylinder placed in a channel using the two-species VCM constitutive model. Secondly, for the first time in numerical simulations, we plan to extend the investigation for two vertically aligned microcylinders placed in a channel for different intercylinder gap ratios, and try to reproduce some of the flow behaviours observed in recent experiments carried out with WLM solutions~\cite{hopkins2020tristability}. Lastly and most importantly, we aim to provide the evidence behind the formation of preferential paths or lanes during the flow of viscoelastic fluids through a porous media based on the analysis of our single and double microcylinders results.

\section{\label{Problem}Problem description and governing equations}
\begin{figure*}
    \centering
    \includegraphics[trim=0cm 0cm 0cm 0cm,clip,width=15cm]{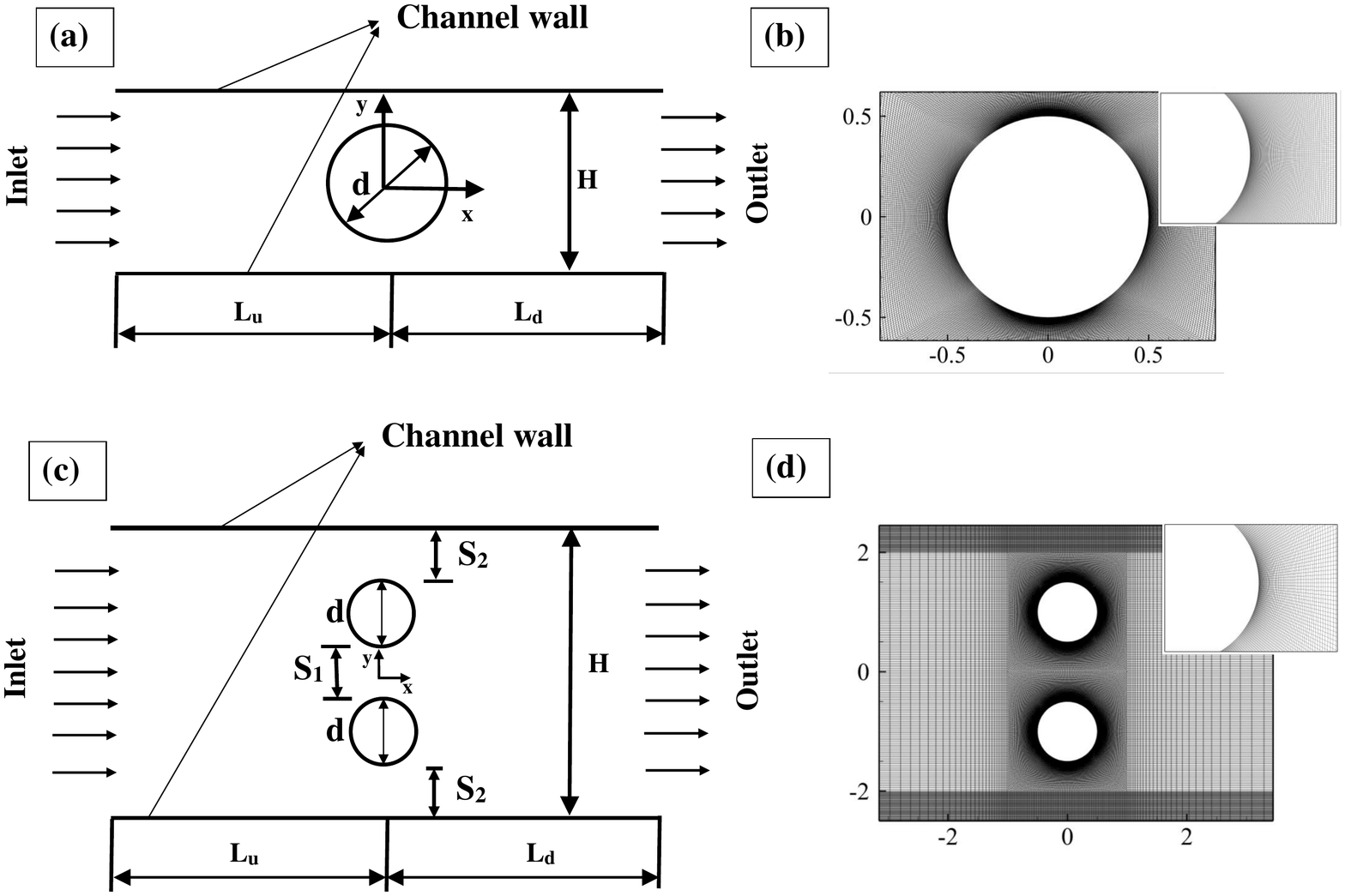}
    \caption{Schematic of the present problem for (a) single microcylinder and (b) side-by-side vertically aligned two microcylinders. Here the flow direction is shown by arrows in the schematic.}
    \label{fig:my_label}
\end{figure*}

 The present study aims to investigate the flow behavior of wormlike micellar solution past a single and two vertically aligned microcylinders of diameter $d$ (or of radius $R$) placed in a rectangular microchannel with different blockage $(BR)$ and gap $(G)$ ratios, as shown schematically in sub Fig.~\ref{fig:my_label}(a) and (c), respectively. The WLM solution enters the channel with a uniform velocity of $U_{in}$. In the case of single cylinder, the blockage ratio is defined as the ratio of the cylinder diameter to that of the channel height, i.e., $BR = \frac{d}{H}$. Whereas, in the case of double cylinders, the gap ratio is defined as $G = \frac{S_{1}}{S_{1} + S_{2}}$, where $S_1$ is the distance between the two cylinders and $S_2$ is the distance between the channel wall and the surface of the cylinder. A value of $G = 0$ implies that the surfaces of the top and bottom cylinders just touch each other, while $G = 1$ indicates that the cylinder surface touches the channel wall. In both the cases, the upstream $(L_{u})$ and downstream $(L_{d})$ length of the channel are kept as $100d$. This length is found to be sufficiently high so that it does not influence the flow dynamics around the microcylinders.      
 
\subsection{Flow equations}
The present flow field will be governed by the following equations, written in their dimensionless forms:\newline
Equation of continuity
\begin{equation}
\label{mass}
    \bm{\nabla} \cdot \bm{U} = 0
\end{equation}
Cauchy momentum equation 
\begin{equation}
\label{mom}
   El^{-1} \frac{D\bm{U}}{Dt} = -\nabla P + \nabla \cdot \bm{\tau}
\end{equation}
In the above equations, $\bm{U}$, $t$ and $\bm{\tau}$ are the velocity vector, time and total extra stress tensor, respectively. All the spatial dimensions are scaled by the cylinder radius $R$, velocity is scaled by $R/\lambda_{eff}$, stress is scaled by the plateau modulus $G_0$ and time is scaled by $\lambda_{eff}$. Here $\lambda_{eff} = \frac{\lambda_{A}}{1+c_{Aeq}^{'}\lambda_{A}}$ is the effective relaxation time for the two-species VCM model in which $\lambda_{A}$ and $c_{Aeq}^{'}$ are the dimensional relaxation time and equilibrium breakage rate of the long worm A, respectively, as discussed in detail in the subsequent subsection. The elasticity number is defined as $El = \frac{Wi}{Re}$, where $Wi = \frac{\lambda_{eff}U_{in}}{R}$ is the Weissenberg number, and $Re = \frac{R U_{in} \rho}{\eta_{0}}$ is the Reynolds number. Here $\rho$ and $\eta_0$ are the solution density and zero-shear rate viscosity, respectively. For an inertialess flow, the left hand side of Eq.~\ref{mom} is essentially zero. The total extra stress tensor, $\bm{\tau}$, for a wormlike micellar solution is given as:
\begin{equation}
    \label{totalStress}
    \bm{\tau} = \bm{\tau_{w}} + \bm{\tau_{s}}  
\end{equation}
where $\bm{\tau_{w}}$ is the non-Newtonian contribution from the wormlike micelles  whereas $\bm{\tau_{s}}$ is the contribution from that of the Newtonian solvent which is equal to $\beta \dot{\bm{\gamma}}$. Here the parameter $\beta$ is the ratio of the solvent viscosity to that of the zero-shear rate viscosity of the wormlike micellar solution and $\dot{\bm{\gamma}} = \nabla \bm{U} + \nabla \bm{U}^{T} $ is the strain-rate tensor. For the two-species VCM model, the total extra stress tensor is given by 
\begin{equation}
     \bm{\tau} = \bm{\tau}_{w} + \bm{\tau_{s}} = (\bm{A} + 2\bm{B})-\left(n_{A} + n_{B}\right)\bm{I}  + \beta\dot{\bm{\gamma}}
\end{equation}
Here $n_{A}$ and $\bm{A}$ are the number density and conformation tensor of the long worm A respectively, whereas $n_{B}$ and $\bm{B}$ are to that of the short worm B. The temporal and spatial evaluation of the number density and conformation tensor for the short and long worms are written in the following subsection based on the VCM model.

 \subsection{Two-species constitutive equations for wormlike micelles: Vasquez-Cook-McKinley (VCM) model}
 
\begin{figure*}
    \centering
    \includegraphics[trim=0cm 0cm 2cm 0cm,clip,width=14cm]{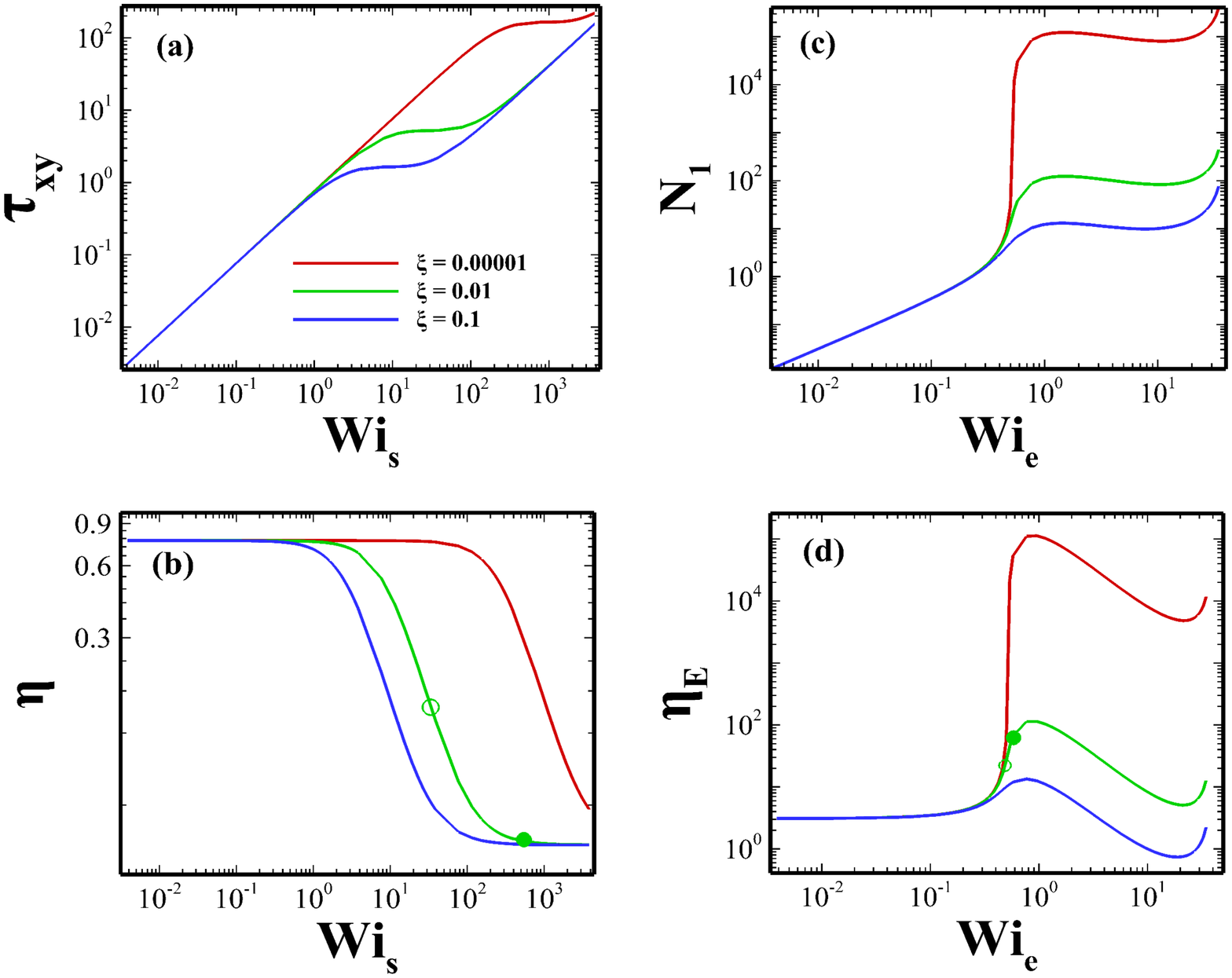}
    \caption{Variations of the non-dimensional shear stress (a) and shear viscosity (b) with the non-dimensional shear rate (or the shear Weissenberg number) and  first normal stress difference (c) and extensional viscosity (d) with the non-dimensional extension rate (or the extensional Weissenberg number) in homogeneous shear and uniaxial extensional flows, respectively. Here the symbols (both filled and open) are used to discuss some results presented in section~\ref{Results}.}
    \label{fig:RheoFlow}
\end{figure*}

 The VCM constitutive equations provide the species conservation equations for the long $(n_{A})$ and short worms $(n_{B})$ along with the equations for the evolution of their conformation tensors $\bm{A}$ and $\bm{B}$, respectively. According to this model, the equations for the variations of $n_{A}$, $n_{B}$, $\bm{A}$, and $\bm{B}$ are given in their non-dimensional forms as follows:
\begin{equation}
    \label{nA}
    \mu\frac{Dn_{A}}{Dt} - 2\delta_{A} \nabla^{2}n_{A} = \frac{1}{2} c_{B} n_{B}^{2} - c_{A}n_{A}
\end{equation}
\begin{equation}
    \label{nB}
    \mu\frac{Dn_{B}}{Dt} - 2\delta_{B} \nabla^{2}n_{B} = - c_{B} n_{B}^{2} + 2 c_{A}n_{A}
\end{equation}
\begin{equation}
    \label{A}
    \mu \bm{A}_{(1)} + A -n_{A} \bm{I} -\delta_{A} \nabla^{2}\bm{A} = c_{B} n_{B} \bm{B} - c_{A} \bm{A}
\end{equation}
\begin{equation}
    \label{B}
    \epsilon \mu \bm{B}_{(1)} + B -\frac{n_{B}}{2} \bm{I} -\epsilon\delta_{B} \nabla^{2}\bm{B} = -2\epsilon c_{B} n_{B} \bm{B} + 2 \epsilon c_{A} \bm{A}
\end{equation}
Here the subscript $( )_{(1)}$ denotes the upper-convected derivative defined as $\frac{\partial()}{\partial t} + \bm{U}\cdot \nabla () - \left( (\nabla \bm{U})^{T} \cdot () + ()\cdot \nabla \bm{U}\right)$. The non-dimensional parameters $\mu$, $\epsilon$ and $\delta_{A,B}$ are defined as $\frac{\lambda_{A}}{\lambda_{eff}}$, $\frac{\lambda_{B}}{\lambda_{A}}$ and $\frac{\lambda_{A} D_{A,B}}{R^{2}}$, respectively, where $\lambda_{B}$ is the relaxation time of the short worm $B$ and $D_{A, B}$ are the dimensional diffusivities of the long and short worms. Furthermore, according to the VCM model, the non-dimensional breakage rate $(c_{A})$ of the long worm A into two equally sized small worms B depends on the local state of the stress field, given by the expression $c_{A} = c_{Aeq} + \mu \frac{\xi}{3}\left( \dot{\bm{\gamma}}: \frac{\bm{A}}{n_{A}} \right)$. On the other hand, the reforming rate of the long worm A from the two short worms B is assumed to be constant, given by the equilibrium reforming rate, i.e., $c_{B} = c_{Beq}$. Here the non-linear parameter $\xi$ is the scission energy required to break a long worm into two equal-sized short worms. The significance of this parameter is that as its value decreases, the amount of stress needed to break a micelle increases. The values of the VCM model parameters chosen for the present study are as follows: $\beta_{VCM} = 10^{-4}$, $\mu = 2.6$, $C_{Aeq} = 1.6$, $C_{Beq} = 0.8607$, $\epsilon = 0.005$, $\delta_{A} = \delta_{B}$ and $\xi = 0.00001, 0.01, 0.1$. The response of the present micellar solution with these VCM model parameters in standard viscometric flows is shown in Fig.~\ref{fig:RheoFlow}. One can see that the solution exhibits the shear-thinning property in shear flows and extensional hardening and subsequent thinning in uniaxial extensional flows, which are very often seen to occur for a wormlike micellar solution.       
Furthermore, one can see that as the value of $\xi$ increases, the shear-thinning tendency of the micellar solution increases, whereas extensional hardening and subsequent thinning tendency decreases.

\section{\label{NumDetail}Numerical details}
A finite volume method based open source computational fluid dynamics code OpenFOAM~\cite{wellerOpenFOAM} and a recently developed rheoFoam solver available in rheotool~\cite{rheoTool} has been used to solve the aforementioned governing equations, namely, mass, momentum, constitutive and number density evaluation equations. All the diffusion terms in the momentum, constitutive and number density equations were discretized using the second-order accurate Gauss linear orthogonal interpolation scheme. All the gradient terms were discretized using the Gauss linear interpolation scheme. While the linear systems of the pressure and velocity fields were solved using the preconditioned conjugate solver (PCG) with DIC (Diagonal-based Incomplete Cholesky) preconditioner, the stress fields were solved using the preconditioned bi-conjugate gradient solver (PBiCG) solver with DILU (Diagonal-based Incomplete LU) preconditioner~\cite{ajiz1984robust,lee2003incomplete}. All the advective terms in the constitutive equations were discretized using the high-resolution CUBISTA (Convergent and Universally Bounded Interpolation Scheme for Treatment of Advection) scheme for its improved iterative convergence properties~\cite{alves2003convergent}. In the present study, the pressure-velocity coupling was established using the SIMPLE (Semi-Implicit Method for Pressure-Linked Equations) method, and the improved both side diffusion (iBSD) technique was used to stabilize the numerical solutions. The absolute tolerance level for the pressure, velocity, stress and micellar concentration fields was set as $10^{-10}$.

A suitable grid density is selected for both the systems by performing the standard grid independence study. In doing so, three different grid densities for each blockage (in the case of single microcylinder) and gap (in the case of two microcylinders) ratio, namely, G1, G2, and G3, consisting of a different number of grid points on the cylinder surface as well as in the whole computational domain were created, and the simulations were run at the highest value of the Weissenberg number considered in the present study. After inspecting the results (in terms of the variation of the velocity, stress and number densities of micelles at different probe locations in the computation domain) obtained for different grid densities, the grid G2 with a range of 59280-82900 (depending upon the blockage ratio) hexahedral cells for the single microcylinder and 83200-88200 (depending upon the gap ratio) hexahedral cells for the two microcylinders cases were found to be adequate for the present study. During the making of any grid, a careful consideration is taken into account. For instance, a very fine mesh is created in the vicinity of the solid cylinder wall to capture the steep gradients of velocity, stress, or concentration fields, whereas a relatively coarse mesh is created away from the solid wall, see sub Figs.~\ref{fig:my_label}(b) and (d). Likewise, the grid independence study, a systematic time independence study was also carried to choose an optimum time step size, and a non-dimensional time step size of 0.00001 was selected for both the systems. The computational domain and its meshing have been done with the help of the blockMeshDict subroutine available in OpenFOAM. Finally, appropriate boundary conditions are employed at different boundaries of the present computational domain to complete the problem description. On the solid surfaces, the standard no-slip and no-penetration boundary conditions for the velocity, i.e., $\bm{U} = 0$ are imposed, whereas a no-flux boundary condition is assumed for both the stress and micellar number density, i.e., $\textbf{n}\cdot \nabla \textbf{A} = 0$ and $\textbf{n}\cdot \nabla \textbf{B} = 0$ and $\textbf{n}\cdot \nabla {n_A} = 0$ and $\textbf{n}\cdot \nabla {n_B} = 0$, where $\textbf{n}$ is the outward unit normal vector. All the simulations were run in a parallel fashion with MPI (Message Passing Interface) interface facility available in OpenFOAM wherein each simulation was distributed among 8 to 12 CPU cores, each of having 2 GB RAM. A detailed validation of the present numerical set up has already been presented in our earlier studies~\cite{sasmal2020flow,khan2020effect}, and hence it is not again performed here.  

\section{\label{Results}Results and discussion}
\subsection{Single microcylinder case : Effect of blockage ratio}
\begin{figure}
    \centering
    \includegraphics[trim=0cm 9cm 0cm 0cm,clip,width=9cm]{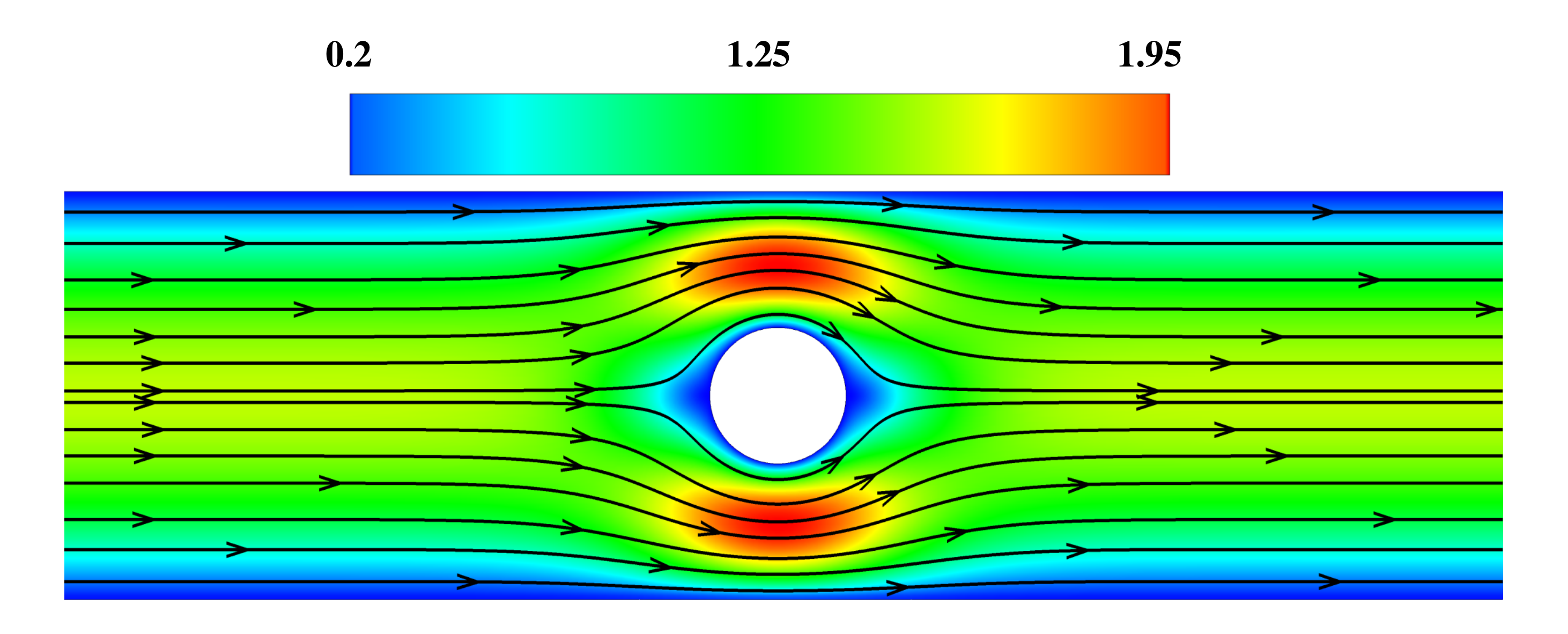}
    \caption{Representative streamline and velocity magnitude plots for Newtonian fluid with blockage ratio of $BR = 0.34$.}
    \label{fig:Newtonian}
\end{figure}

Before studying the complex flow dynamics of a wormlike micellar solution, first, we present the results of the flow behavior of a simple Newtonian fluid around a single microcylinder confined between two parallel channel walls at different blockage ratios. Figure~\ref{fig:Newtonian} shows the streamlines and velocity magnitude plots of a Newtonian fluid at a particular value of $BR = 0.34$. It can be clearly seen that both the streamline and velocity magnitude plot show a perfect fore-aft symmetry along the horizontal and vertical mid planes passing through the origin, as expected for a simple Newtonian fluid flowing under the creeping flow condition. The streamlines just follow a smooth order and steady path without crossing to each other. Furthermore, the streamlines are seen to be attached with the cylinder surface and hence, no separation of flow happens. This result is inline with that observed in our earlier numerical study~\cite{khan2020effect} and experimental observation of Zhao et. al.~\cite{zhao2016flow}. The velocity magnitude is seen to be maximum in the narrow gap between the channel wall and cylinder surface. For other blockage ratios considered in this study, a similar flow pattern is observed for the Newtonian fluid. The only difference seen is that the maximum velocity magnitude in the gaps between the channel wall and cylinder surface decreases as the blockage ratio decreases. This is simply due to an increase in the flow area with the decreasing value of the blockage ratio.          

\begin{figure}
    \centering
    \includegraphics[trim=0cm 6.5cm 0cm 0cm,clip,width=9cm]{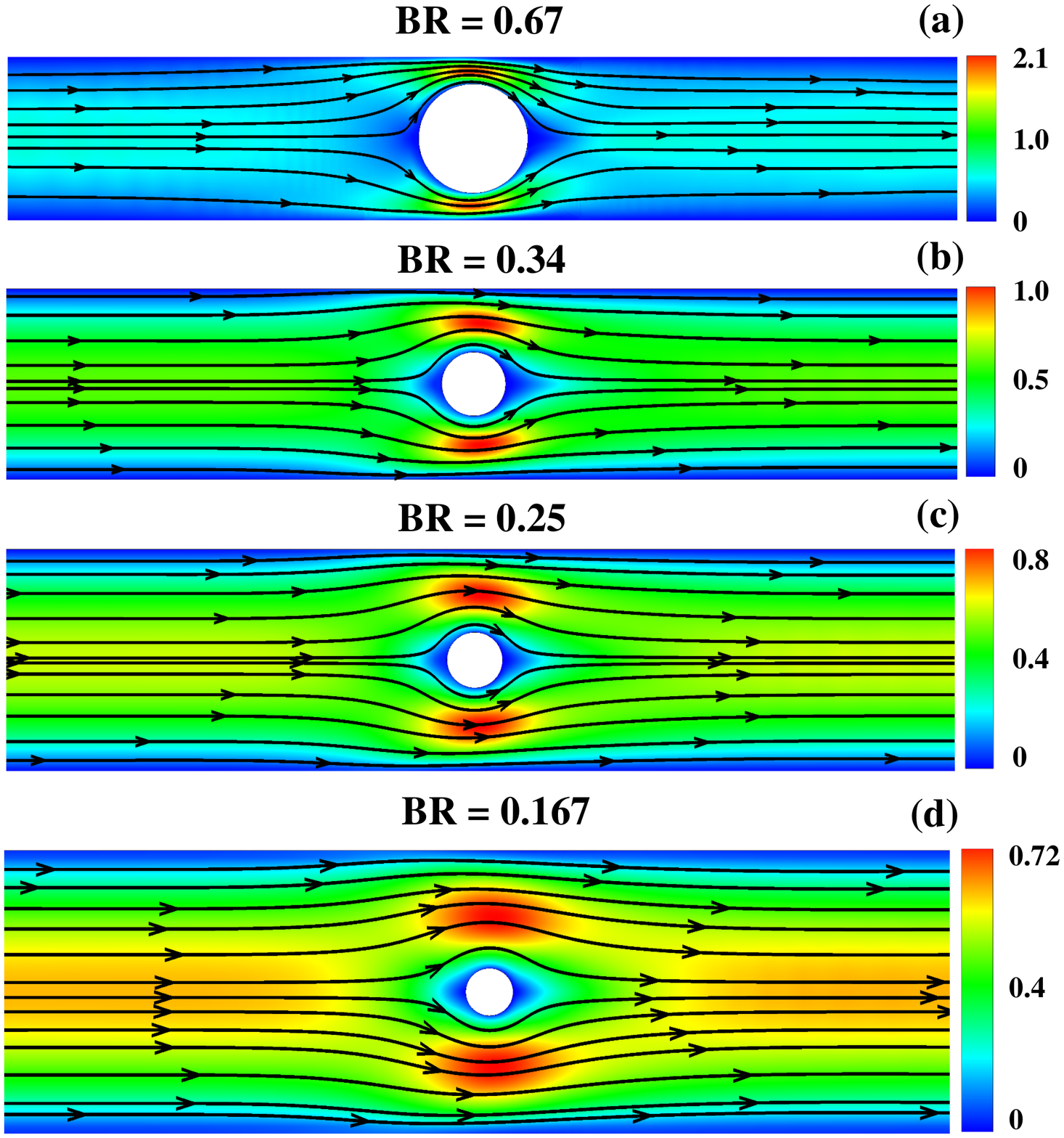}
    \caption{Representative streamline and velocity magnitude plots of a WLM solution at $Wi = 1.0$ and $\xi = 0.01$ for different blockage ratios.}
    \label{fig:streamline_wi1.0}
\end{figure}

\begin{figure}
    \centering
    \includegraphics[trim=0cm 0cm 0cm 0cm,clip,width=13cm]{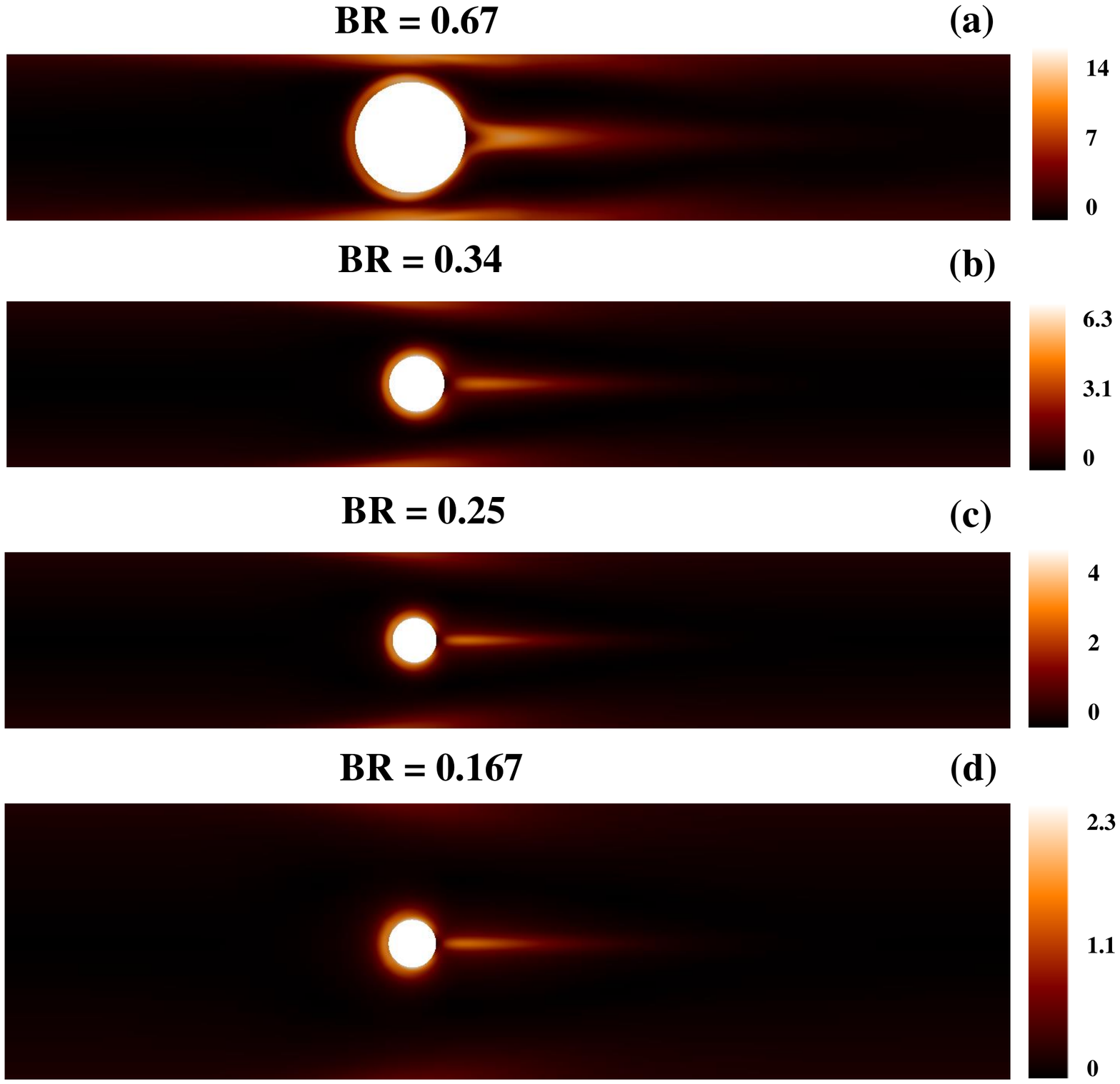}
    \caption{Surface plot of principle stress difference of a WLM solution at $Wi = 1.0$ and $\xi = 0.01$ for different blockage ratios.}
    \label{fig:psd_wi1.0}
\end{figure}

Unlike the Newtonian fluid, the flow of WLM solutions is expected to be strongly dependent on the blockage ratio due to its complex rheological behaviour. Additionally, one can expect a strong dependency on the values of the non-dimensional parameters like the Weissenberg number and non-linear VCM model parameter $\xi$. At very low values of the Weissenberg number, for instance at $Wi = 0.01$, the flow behaviour of WLM solutions at different blockage ratios is found to be similar as that observed for the Newtonian fluid (results are not shown here). This is due to the presence of a weak viscoelastic effect. However, as the Weissenberg number gradually increases to higher values, the flow dynamics become strongly dependent on the values of the blockage ratio, Weissenberg number and non-linear VCM model parameter $\xi$. As for example, at $Wi = 1$, although the flow remains steady, and the streamlines follow a nice order path as that seen for Newtonian fluid and WLM solutions at $Wi = 0.01$, the symmetry in the flow profiles along the vertical mid-plane passing through the origin starts to break, Fig.~\ref{fig:streamline_wi1.0}. As the blockage ratio increases, the tendency of destroying this vertical symmetry increases, for instance, see the results in sub Figs~\ref{fig:streamline_wi1.0}(b) and (d) at the values of $BR = 0.34$ and 0.167, respectively. However, the horizontal symmetry still exists at this value of the Weissenberg number irrespective of the value of $BR$. The corresponding surface plot of the non-dimensional principle stress difference, defined as $PSD = \sqrt{\left( \tau_{xx} - \tau_{yy}\right)^{2} + \left( 2 \tau_{xy}\right)^{2}}$, is presented in Fig.~\ref{fig:psd_wi1.0} at different blockage ratios. Regardless of the blockage ratio, the PSD value is seen to be high in the vicinity of the cylinder surface due to the presence of a high shearing zone. Apart from this, a strand of high PSD value, also known as the birefringent strand, is formed along the mid horizontal plane downstream of the cylinder. This is due to the formation of a highly extensional flow field in this region, which thereby aligning more long micelles in the flow field as well as breaking them into smaller ones. Both these facts tend to increase the PSD value in this region. As the blockage ratio increases, the thickness as well as the value of this birefringent strand increases due to an increase both in the shear and extensional flow strengths.  

\begin{figure}
    \centering
    \includegraphics[trim=0cm 1cm 0cm 0cm,clip,width=9cm]{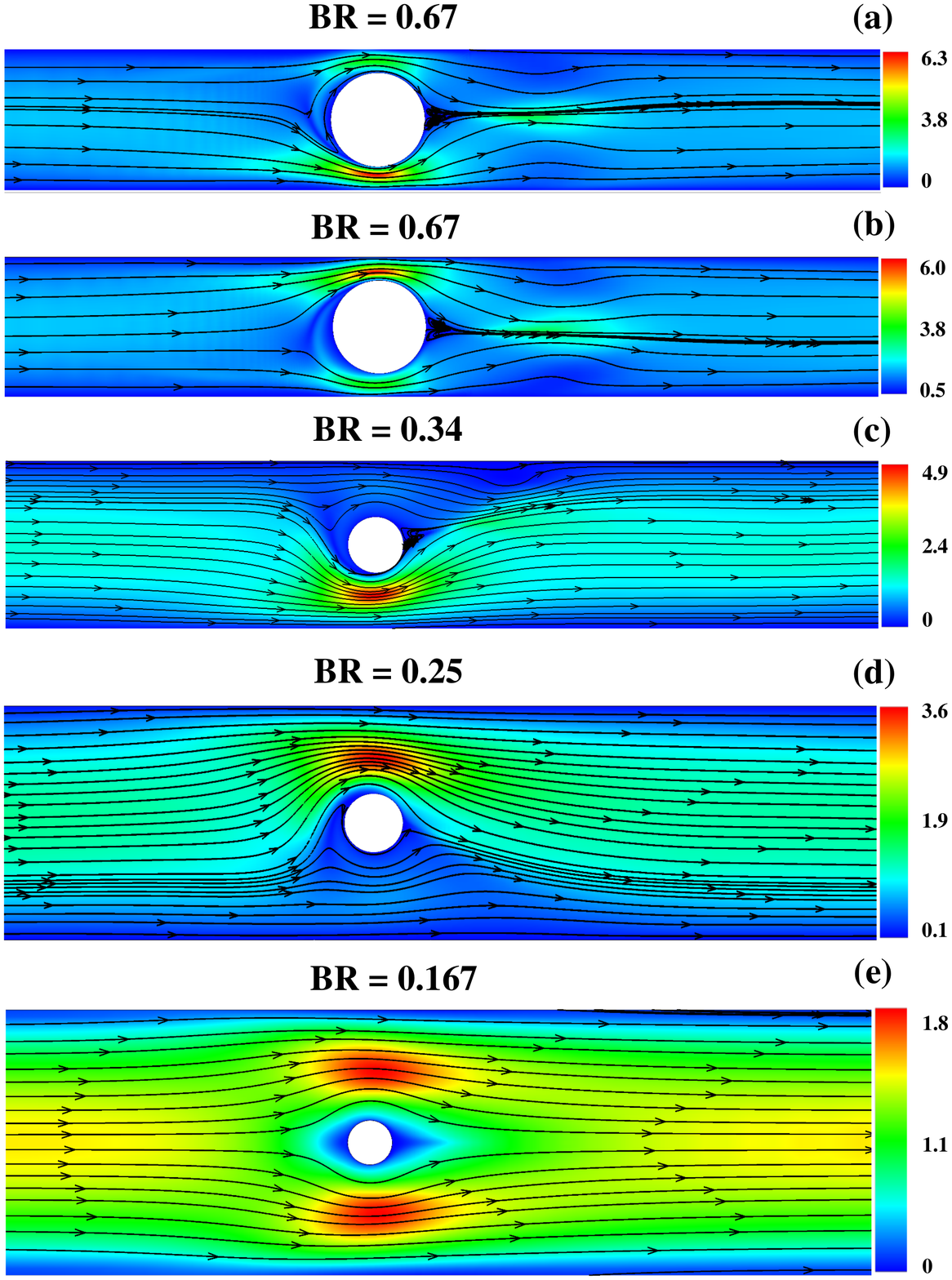}
    \caption{Representative streamline and velocity magnitude plots of a WLM solution at $Wi = 2.5$ and $\xi = 0.01$ for different blockage ratios.}
    \label{fig:streamline_wi2.5}
\end{figure}

\begin{figure*}
    \centering
    \includegraphics[trim=0cm 0cm 0cm 0cm,clip,width=15cm]{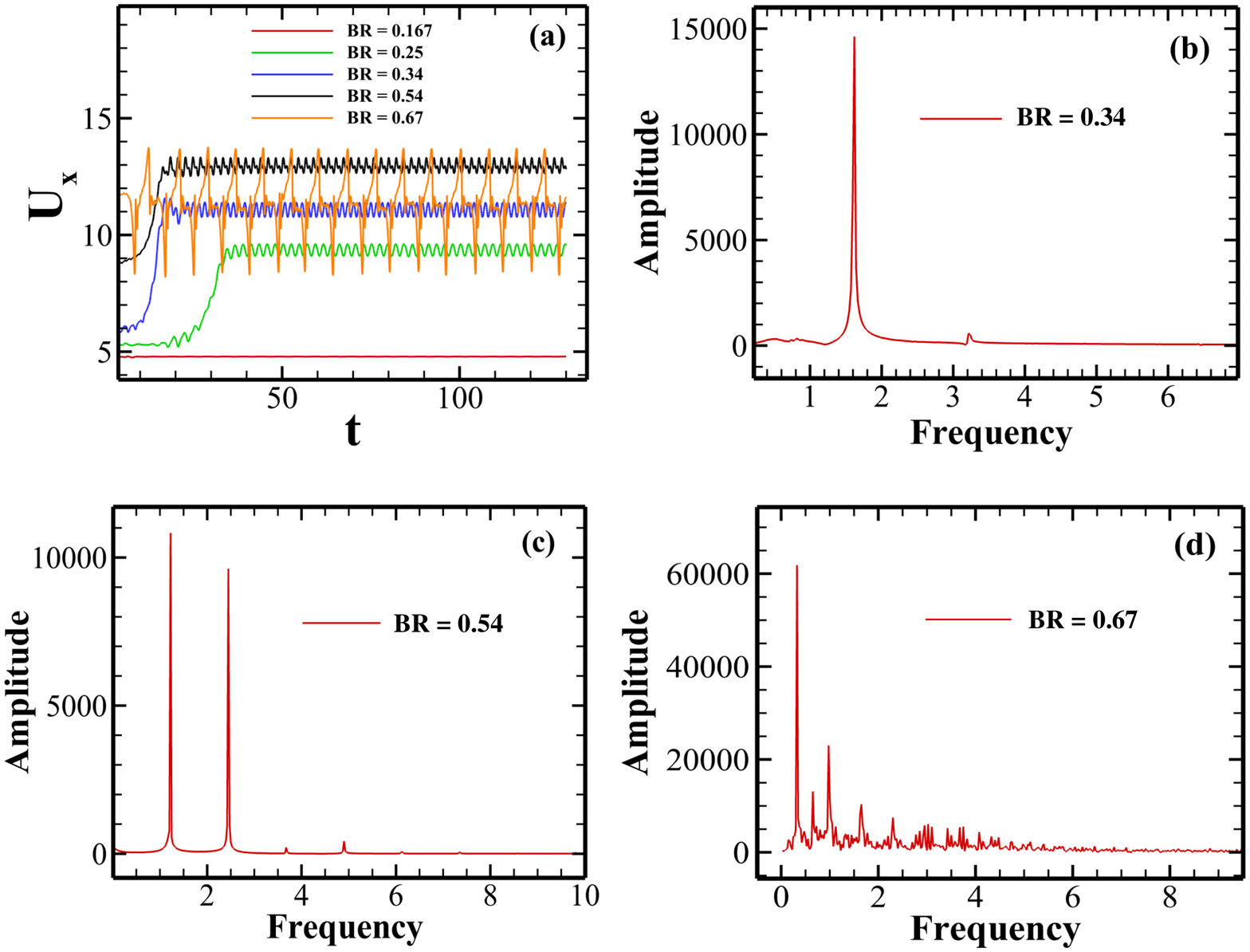}
    \caption{(a) Temporal variation of the stream-wise velocity component at a probe location ...  and (b-d) power spectral density plot of the velocity fluctuations at different blockage ratios at $Wi = 2.5$ and $\xi = 0.01$.}
    \label{fig:temp_fft}
\end{figure*}

As the value of the Weissenberg number is further incremented, say to 2.5, the flow remains steady and horizontally symmetric in the case of the least blockage ratio of $BR = 0.167$, sub Fig.~\ref{fig:streamline_wi2.5}(e). On the other hand, at the maximum blockage ratio of $BR = 0.67$ considered in this study, the flow becomes unsteady and quasi-periodic at the same Weissenberg number. At this blockage ratio, a distortion in the streamline profiles is observed, particularly at the rear side of the cylinder. Furthermore, the region of the maximum velocity magnitude changes its position between the lower (sub Fig.~\ref{fig:streamline_wi2.5}(a)) and upper narrow gap (sub Fig.~\ref{fig:streamline_wi2.5}(b)) regions situated in between the channel wall and cylinder surface. This suggests the emergence of an elastic instability in the flow field, and an elastic wave downstream of the cylinder due to the shifting in the maximum velocity magnitude zone between the two gap regions, as discussed and explained in detail in our earlier study~\cite{khan2020effect}. Moreover, a small vortex is seen to form downstream of the cylinder at this blockage ratio and Weissenberg number. The nature of the flow field at these two extreme blockage ratios, namely, at $BR = 0.167$ and 0.67, is further confirmed in Fig.~\ref{fig:temp_fft}(a) wherein the temporal variation of the non-dimensional stream-wise velocity is plotted at a probe location placed at the mid-point in between the cylinder surface and channel wall for different blockage ratios. At $BR = 0.167$, it reaches to a steady value with time, suggesting the presence of a steady state flow field. Whereas, at $BR = 0.67$, it fluctuates with time and therefore shows the occurrence of unsteadiness in the flow field. The power spectrum of these velocity fluctuations is presented in sub Fig.~\ref{fig:temp_fft}(d), and from this figure, it can be seen that the flow is governed by a single dominant frequency along with a broad spectrum of small frequencies. This indicates the quasi-periodic nature of the flow field at these values of $Wi$ and $BR$. 

\begin{figure}
    \centering
    \includegraphics[trim=0.5cm 2cm 0cm 0cm,clip,width=13cm]{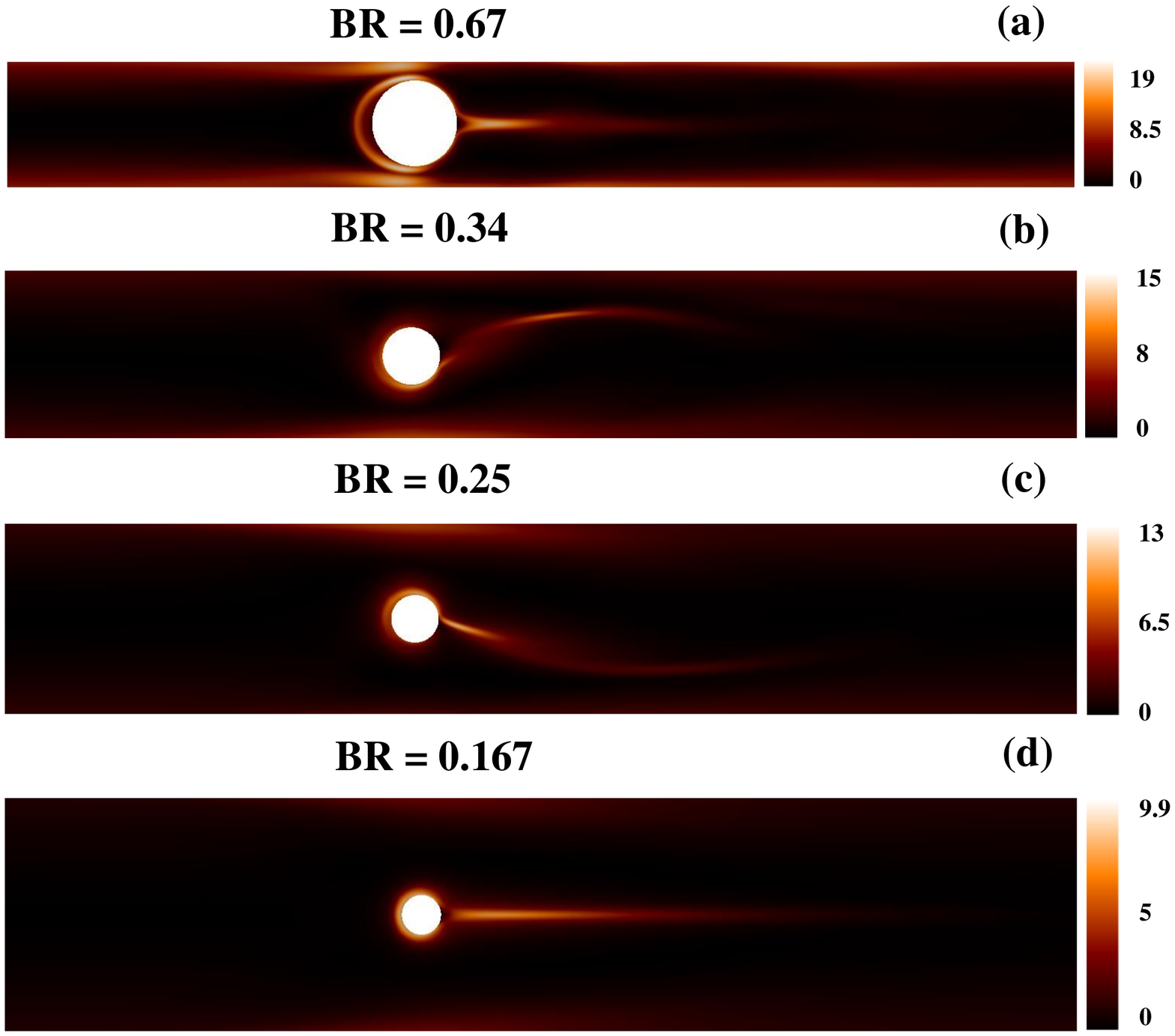}
    \caption{Surface plot of principle stress difference of a WLM solution at $Wi = 2.5$ and $\xi = 0.01$ for different blockage ratios.}
    \label{fig:psd_wi2.5}
\end{figure}

In between these two extreme blockage ratios considered in this study, there is a range of blockage ratio present wherein the fluid prefers to flow through one side of the cylinder, for instance, see sub Figs.~\ref{fig:streamline_wi2.5}(c) and (d) for the results at $BR = 0.34$ and 0.25, respectively. This results in the formation of an almost stagnant region on the opposite side of the cylinder.  Here the preferential side occurs at $Y < 0$ for $BR = 0.34$ (sub Fig.~\ref{fig:streamline_wi2.5}(c)), whereas for $BR = 0.25$, it occurs at $Y > 0$ (sub Fig.~\ref{fig:streamline_wi2.5}(d)). However, the selection of this preferential side for the flow is completely random, and hence, there is an equal opportunity present when the fluid can go through the other side of the cylinder. The occurrence of this flow asymmetry indicates the origin of a pitchfork bifurcation in the flow field. This kind of bifurcation in the flow field has also been observed in earlier experimental investigations dealing with polymer~\cite{haward2020asymmetric} and WLM solutions~\cite{haward2019flow}, as well as in numerical investigations performed with a single-species viscoelastic constitutive model~\cite{varchanis2020asymmetric}. At $BR = 0.34$, the flow field seems to be unsteady in nature, whereas it is steady at $BR = 0.25$, which can be seen from the temporal variation of the non-dimensional stream-wise velocity presented in sub Fig.~\ref{fig:temp_fft}(a). The corresponding power spectrum plot for velocity fluctuations at $BR = 0.34$ is depicted in sub Figs.~\ref{fig:temp_fft}(b). From this figure, one can see that the flow is governed by a single dominant frequency, thereby suggesting the occurrence of a regular periodic unsteadiness in the flow field. At $BR = 0.57$, an asymmetry in the flow field is also seen (results not shown here), and the flow field is again found to be unsteady, which is quasi-periodic in nature as can be evident from the power spectrum plot of velocity fluctuations presented in sub Fig.~\ref{fig:temp_fft}(c).The corresponding variation of the PSD value at $Wi = 2.5$ and at different blockage ratios is depicted in Fig.~\ref{fig:psd_wi2.5}. Once again, at this Weissenberg number, a long birefringent strand of high PSD value is seen to form downstream of the cylinder likewise it is seen at $Wi = 1$ (Fig~\ref{fig:psd_wi1.0}). However, the PSD value is higher at $Wi = 2.5$ than that seen at $Wi = 1$ due to an increase in the flow strength. Furthermore, the strand is seen to be bending in nature downstream of the cylinder at blockage ratios 0.34 (sub Fig.~\ref{fig:psd_wi2.5}(b)) and 0.25 (sub Fig.~\ref{fig:psd_wi2.5}(c)) due to the presence of an asymmetric flow at these blockage ratios.   

\begin{figure}
    \centering
    \includegraphics[trim=1.3cm 0cm 0cm 0cm,clip,width=9.5cm]{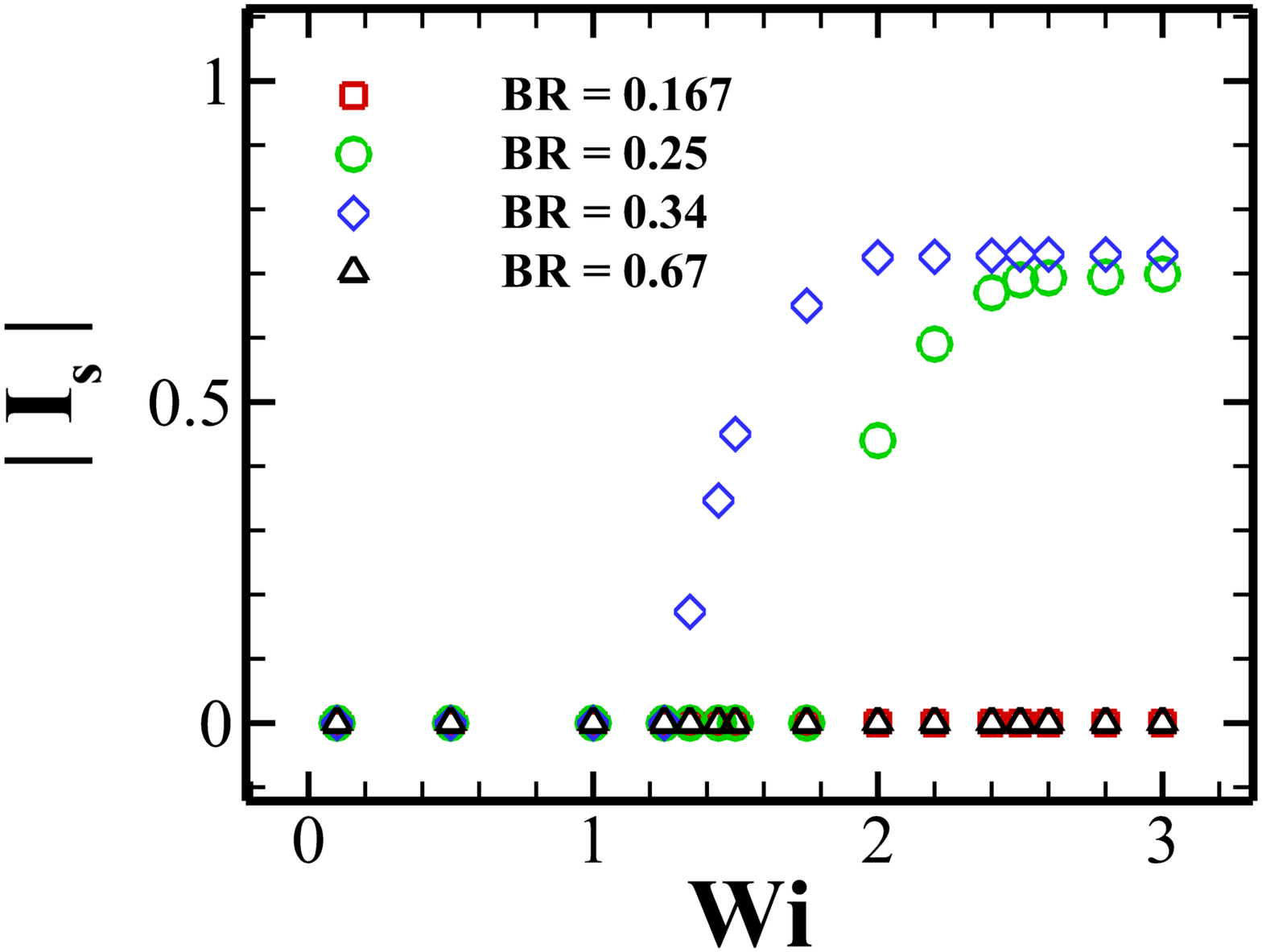}
    \caption{Variation of the flow asymmetry parameter $(I_{s})$ with the Weissenberg number and blockage ratio at $\xi = 0.01$.}
    \label{fig:IvsWi}
\end{figure}

To characterize the asymmetric nature of the flow more quantitatively, we define a dimensionless flow asymmetry parameter $I_{s}$ as follows~\cite{varchanis2020asymmetric,haward2019flow}
\begin{equation}
    I_{s} = \frac{U_{X,1}-U_{X,2}}{U_{X,1}+U_{X,2}}
\end{equation}
Here $U_{X,1}$ and $U_{X,2}$ are the stream-wise velocities at the midpoints in between the cylinder surface and upper and lower channel walls, respectively. A value of $|I_{s}| = 0$ denotes a perfect symmetric flow; whereas, $|I_{s}| = \pm 1$ implies a perfect asymmetric flow when the whole fluid passes through one side of the cylinder. Note that in the case of an unsteady flow, a time averaged value of $U_{X}$ is considered in the calculation of $I_{s}$. The variation of the absolute value of $I_{s}$ with the Weissenberg number and blockage ratio is presented in Fig~\ref{fig:IvsWi}. It can be seen that the value of $I_{s}$ is essentially zero for the blockage ratios of 0.17 and 0.67. This is due to the existence of the steady symmetric and unsteady symmetric quasi-periodic flows at these two blockage ratios, respectively. On the other hand, at blockage ratios 0.25 and 0.34, a critical value of the Weissenberg number is seen to present up to which the asymmetry parameter is zero, and beyond that it suddenly starts to increase and finally reaches almost to a constant value at high Weissenberg numbers. The critical value of the Weissenberg number at which the transition from symmetric to an asymmetric flow occurs (i.e., the onset of the pitchfork bifurcation), increases as the blockage ratio decreases. For instance, at $BR = 0.34$, it is around 1.25 while it is around 1.75 at $BR = 0.25$.   

\begin{figure}
    \centering
    \includegraphics[trim=0cm 0cm 0cm 0cm,clip,width=9.3cm]{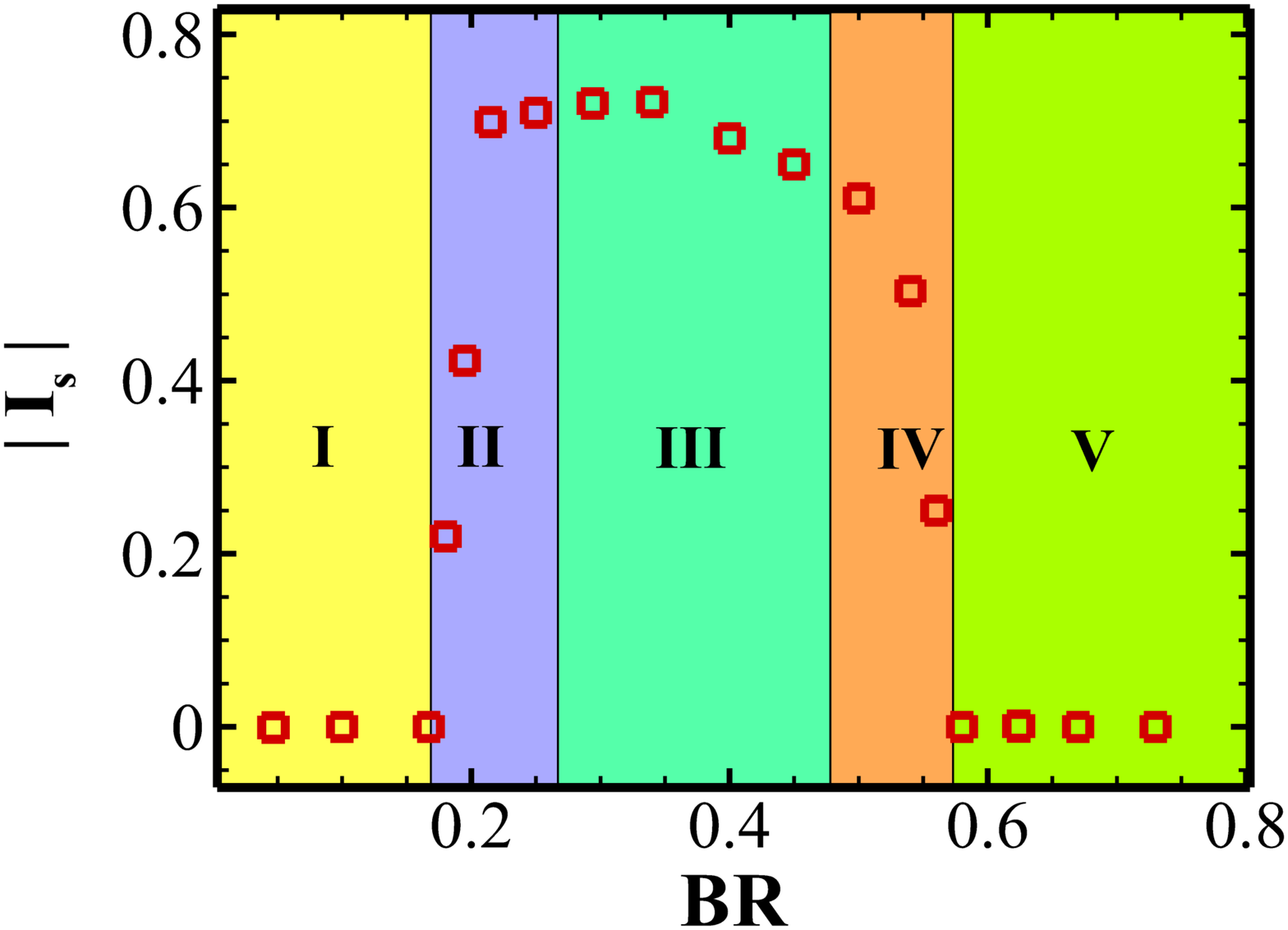}
    \caption{Variation of the flow asymmetry parameter $(I_{s})$ with the blockage ratio at $Wi = 2.5$ and $\xi = 0.01$. In this figure (I) steady and symmetric (II) steady and asymmetric (III) unsteady, periodic and asymmetric (IV) unsteady, quasi-periodic and asymmetric and (V) unsteady, quasi-periodic and symmetric.}
    \label{fig:IvsBR}
\end{figure}
Furthermore, one can see that the value of the flow asymmetry parameter $I_{s}$ increases with the blockage ratio, which is in line with that observed by Varchanis et al.~\cite{varchanis2020asymmetric} in their simulations. Based on the value of the flow asymmetry parameter, a phase diagram is presented in Fig.~\ref{fig:IvsBR} wherein different flow states observed in the present study with the blockage ratio, are summarized at a Weissenberg number of 2.5 and non-linear VCM model parameter $\xi = 0.01$. At a blockage ratio lower than 0.167, the flow is steady and symmetric. Beyond that and up to $BR = 0.27$, a transition to a steady and asymmetric flow occurs. After that the flow transits to an unsteady periodic state and then to a quasi-periodic state as the blockage ratio gradually increases. On further increasing the blockage ratio of more than around 0.55, the flow transits to a quasi-periodic and symmetric state where a resymmetrization in the flow occurs.

Next, we aim to explain the origin of this asymmetric flow resulting from the flow bifurcation and elastic instabilities in WLM solutions. It is well known that the onset of elastic instabilities either in polymer or micellar solutions is the resultant of the presence of curved streamlines in the vicinity of the microcylinder and the accumulation of the elastic stresses downstream of the microcylinder~\cite{pakdel1996elastic,mckinley1996rheological,fardin2012instabilities,zhao2016flow}, which can be seen from the streamlines plot (Fig.~\ref{fig:streamline_wi2.5}) and the PSD contours (Fig.~\ref{fig:psd_wi1.0}) presented here as well. Very often, the criteria developed by McKinley and co-workers are used to figure out the onset of these purely elastic instabilities, written as~\cite{mckinley1996rheological}
\begin{equation}
    \left( \frac{\lambda U}{\mathscr{R}} \frac{\tau_{xx}}{\eta_{0} \dot{\gamma}} \right) \geq M_{crit}^{2}
    \label{Mcrit}
\end{equation}
where $\mathscr{R}$ is the characteristic radius of streamline curvature and $\tau_{xx}$ is the tensile or normal stress along the flow direction. If the dimensionless value of the left hand side of Eq.~\ref{Mcrit} becomes greater than or equal to the critical $M_{crit}^{2}$ value at any position in the flow field, an instability will then be originated in the system. For the flow of a constant viscosity viscoelastic polymer (Boger fluid) solution past a cylinder confined in a channel, a value of $M_{crit} = 6.08$ was found from the linear stability analysis~\cite{mckinley1996rheological}. However, for the present case of a wormlike micellar solution, this value should not be obviously the same due to the presence of shear-thinning viscous properties and breakage and reformation dynamics of the micelles. Once this instability is triggered in the flow field, then a small and random lateral fluctuation of the birefringent strand (as shown in Fig.~\ref{fig:psd_wi2.5}) of high elastic stresses downstream of the cylinder either in the $-Y$ or $+Y$ direction creates a resistance to the flow of fluid in that direction. This forces the fluid to pass through the other side of the cylinder. This will eventually create an imbalance in the shear rate at the two sides of the cylinder. If the fluid shows shear-thinning properties, this imbalance in the shear rate and hence the viscosity gets accentuated, thereby resulting in the fluid to pass through one side (at which the shear rate is high or the viscosity is low) of the cylinder. This explanation is in line with that provided earlier for the flow of either WLM solution~\cite{haward2019flow} or polymer solution~\cite{haward2020asymmetric} past a cylinder. Therefore, to show the asymmetric flow, the fluid should have shear-thinning properties and a sufficient amount of elastic stresses should be accumulated downstream of the cylinder~\cite{haward2020asymmetric}. 

To explicitly explain this, we calculate the local shear $(Wi_{s}^{l})$ and extensional $(Wi_{e}^{l})$ Weissenberg numbers based on the local shear rate in the gap region and local extension rate downstream of the cylinder respectively for $BR = 0.34$, $Wi = 2.5$ and $\xi = 0.01$ at which an asymmetric flow was observed (sub Fig.~\ref{fig:streamline_wi2.5}(c)). We find that these values (presented as open symbols in Fig.~\ref{fig:RheoFlow}) are lied in the shear-thinning region (in case of the shear Weissenberg number) and extensional hardening region (in case of the extensional Weissenberg number) in the plots presented in Fig.~\ref{fig:RheoFlow}. As the blockage ratio increases to 0.67, the values (presented as filled symbols in Fig.~\ref{fig:RheoFlow}) of both $(Wi_{s}^{l})$ and  $(Wi_{e}^{l})$ increase due to the increase in the flow velocity resulting from the decrease in the flow area. Once again, these values are shown in the same figure as symbols, and one can see that although the value of $(Wi_{e}^{l})$ lies in the extensional hardening region, the value of $(Wi_{s}^{l})$ lies in the plateau region in shear viscosity plot. This causes a resymmetrization in the flow field at this blockage ratio as shown in sub Figs.~\ref{fig:streamline_wi2.5} (a) and (b). 

\begin{figure}
    \centering
    \includegraphics[trim=0.5cm 0.8cm 0cm 0cm,clip,width=9.5cm]{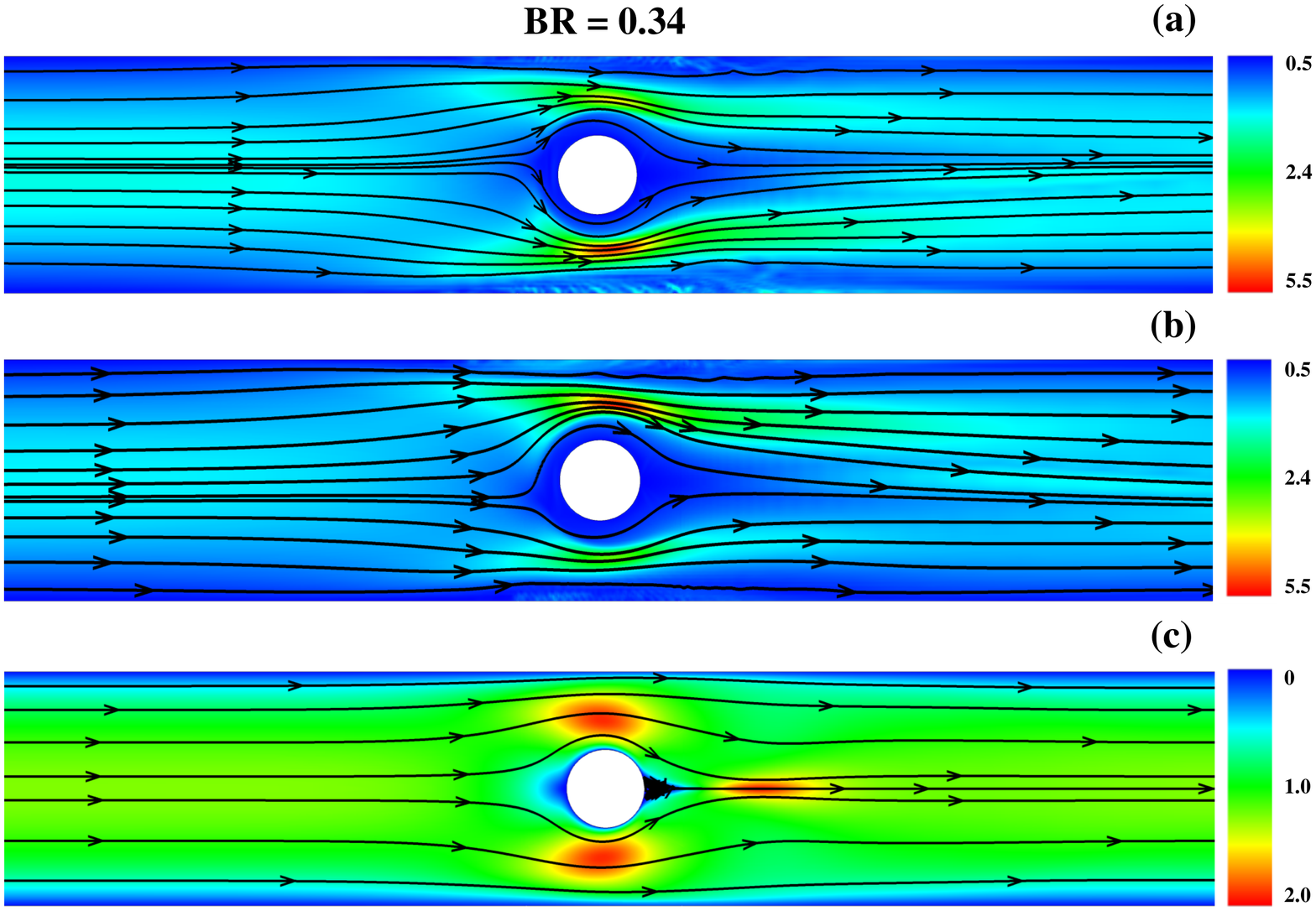}
    \caption{Representative streamline and velocity magnitude plots at $BR = 0.34$ and $Wi = 2.5$. (a) and (b) $\xi = 0.00001$, (c) $\xi = 0.1$.}
    \label{fig:stream_Xi0.0001}
\end{figure}

This is further confirmed by changing the value of $\xi$ which indicates the scission energy needed to break a micelle. As the value of $\xi$ increases to 0.1 or the micelles become progressively easier to break, a symmetric flow (with $|I_{s}| = 0$) is seen to present (sub Fig.~\ref{fig:stream_Xi0.0001}(c)) at the same $BR = 0.34$ and $Wi = 2.5$ as opposed to an symmetric flow seen at $\xi = 0.01$. 
This is simply due to the fact that although the shear-thinning property increases with an increase in $\xi$ due to an easy breakage of micelles, the magnitude of the elastic stresses downstream of the cylinder becomes insufficient to create instability in the system. On the other hand, further simulations were also run to a lower value of $\xi = 0.0001$ at which the micelles become more harder to break. It can be again seen a resymmetrization in the flow field, sub Figs.~\ref{fig:stream_Xi0.0001}(a) and (b) shown at two different times. At this value of $\xi$, although the value of $Wi_{e}^{l}$ increases, the value of $Wi_{s}^{l}$ lies in the plateau region shown in Fig.~\ref{fig:RheoFlow}.

\subsection{Two vertically aligned microcylinders case: Effect of gap ratio}
\begin{figure*}
    \centering
    \includegraphics[trim=1cm 5cm 0cm 0cm,clip,width=18cm]{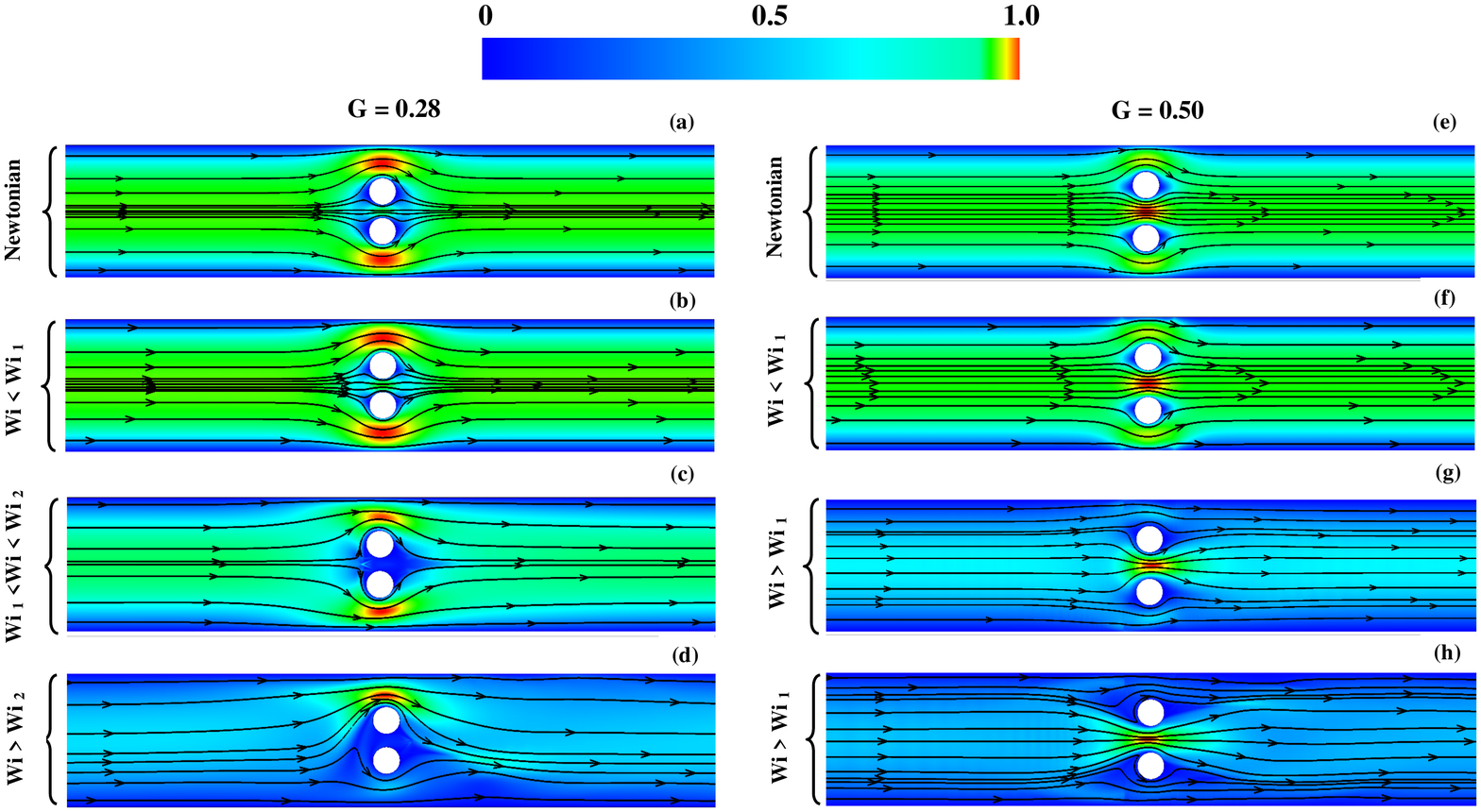}
    \caption{Representative streamline and velocity magnitude plots for vertically side-by-side two microcylinders case at $\xi = 0.01$.}
    \label{fig:sidebyside_velmag}
\end{figure*}

After discussing the results for the case of a single microcylinder, we now turn our attention to the present and discuss the results for two vertically side-by-side placed microcylinders in a channel, as schematically shown in Fig.~\ref{fig:my_label}(c). The streamlines and velocity magnitude plots for this configuration are depicted in Fig.~\ref{fig:sidebyside_velmag} at two gap ratios, namely, 0.28 (a-d) and 0.50 (e-f) for a range of values of the Weissenberg number. Likewise the single cylinder case, for a Newtonian fluid, a perfect symmetry along the horizontal and vertical mid-planes passing through the origin, is present in the flow profiles irrespective of the value of the gap ratio $G$, see sub Fig~\ref{fig:sidebyside_velmag}(a) and (e). Although the fluid passes through all the three gaps available in the system; however, at $G = 0.28$, the magnitude of the velocity is larger at the gap regions in between either the top or bottom cylinder and the channel wall than that seen at the gap region in between the two cylinders. In contrast to this, a reverse trend is seen for the gap ratio of $G = 0.50$. This is simply due to the fact that for a Newtonian fluid and in the creeping flow regime, the volumetric flow rate of the fluid is linearly proportional to the available flow area. At G = 0.28, the flow area is larger at the gap in between either the top or bottom cylinder and the channel wall than that seen in between the two cylinders; whereas, at $G = 0.50$, the other way around happens. Below a critical low value of the Weissenberg number $ Wi < Wi_{1} \approx 0.3$, the flow characteristics of a WLM solution look similar to that of a Newtonian fluid regardless of the gap ratio, as it was also seen for the single cylinder case. For instance, see the results that are presented in sub Fig.~\ref{fig:sidebyside_velmag}(b) and (f) for gap ratios of 0.28 and 0.50, respectively. This is solely due to the fact that at this low Weissenberg and Reynolds number flows, the elastic effects as well as the breakage and reformation dynamics of micelles are very weak and hence, it behaves like a Newtonian fluid.

\begin{figure*}
    \centering
    \includegraphics[trim=0cm 0cm 1cm 0cm,clip,width=14cm]{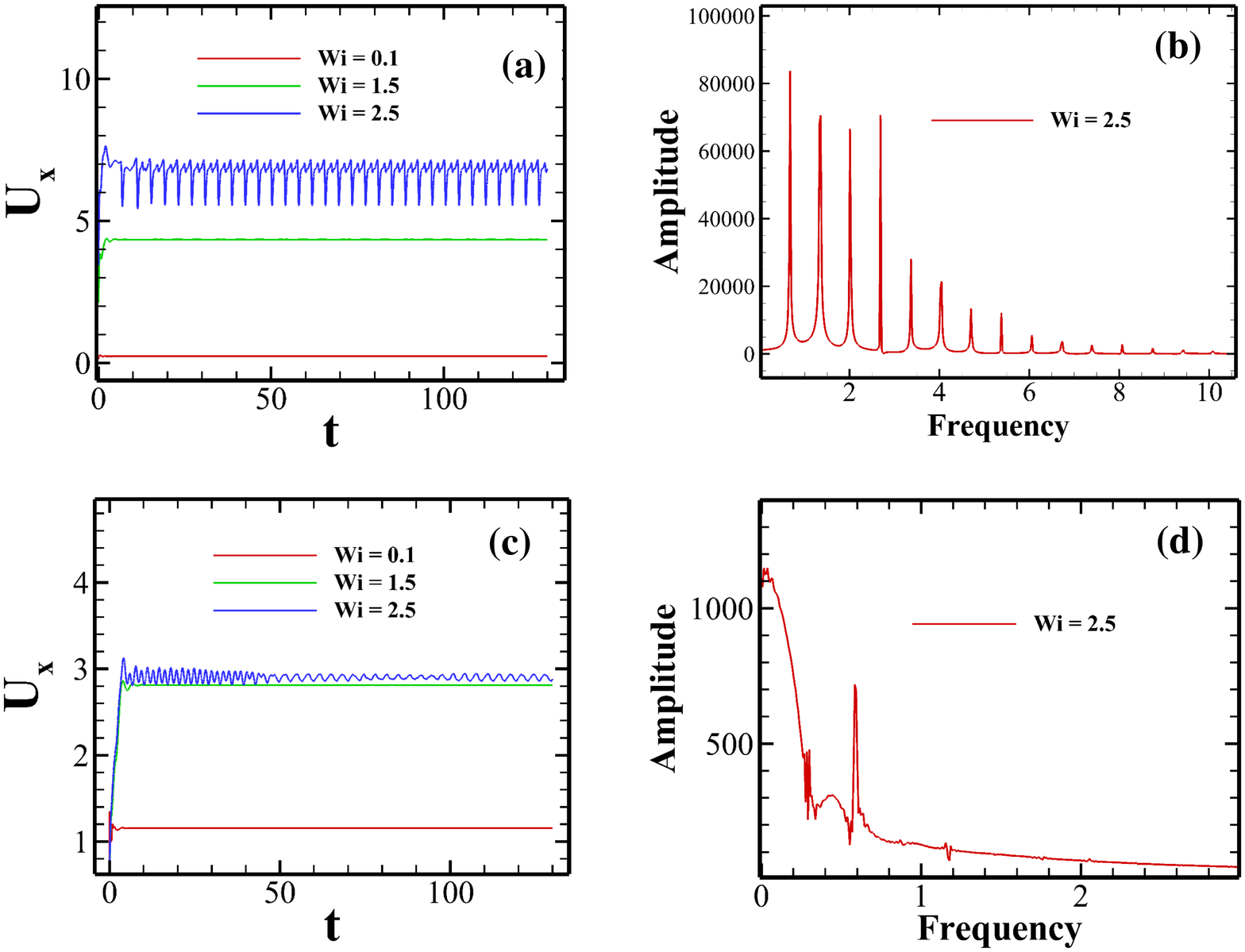}
    \caption{Temporal variation of the stream-wise velocity component at a probe location $X = 0$ and $Y = 0$ for two gap ratios, namely, 0.28 (a) and 0.5 (b). The corresponding power spectral density plot of the velocity fluctuations at $G = 0.28$ (b) and at $G = 0.5$. Here all the results are presented for non-linear VCM model parameter $\xi = 0.01$.}
    \label{fig:velvstime}
\end{figure*}

However, as the Weissenberg number gradually increases to higher values and exceeds the first critical Weissenberg number $(Wi_{1})$, the system then undergoes the first transition due to the increase in the elastic forces. For instance, at $G = 0.28$, a transition from the low-Weissenberg number symmetric state to a diverging state (D) state occurs, in which the fluid passes through the gaps in between the cylinder and channel wall, and it completely avoids the region in between the two cylinders, sub Fig.~\ref{fig:sidebyside_velmag}(c). The flow still remains steady and symmetric along the horizontal mid-plane passing through the origin, as can be observed in sub Fig.~\ref{fig:velvstime}(a), wherein the temporal variation of the non-dimensional stream-wise velocity is plotted at a probe location placed at the origin. On further increasing the Weissenberg number beyond a second critical value of the Weissenberg number $Wi > Wi_{2}$, a second transition in the flow state is observed, in which the micellar solution mostly prefers to flow through only the gap in between the top cylinder and the channel wall $(Y > 0)$, as shown in sub Fig.~\ref{fig:sidebyside_velmag}(d). However, there is an equal opportunity present in which most of the fluid can also pass through the gap in between the bottom cylinder and the channel wall $(Y < 0)$ (not shown here). This state is known as the asymmetric-diverging state (AD). In this state, the flow becomes unsteady, as can be evident in sub Fig.~\ref{fig:velvstime}(a) wherein the non-dimensional stream-wise velocity is seen to be fluctuating with time. The nature of this unsteadiness is quasi-periodic as the power-spectrum of the velocity fluctuations is governed by more than one dominant frequencies, sub Fig.~\ref{fig:velvstime}(b). This state is analogous to the state observed in sub Fig.~\ref{fig:streamline_wi2.5}(d) for the case of a single cylinder. On the other hand, at $G = 0.5$, only one transition in the flow state happens when the Weissenberg number exceeds its first critical value $Wi > Wi_{1}$. In this state, the whole micellar solution preferentially passes through the gap region in between the two cylinders, avoiding the gap in between the cylinder and the channel wall. This state is known as the converging state (C). However, a transition from a steady flow field to an unsteady one occurs within this state as the Weissenberg number gradually increases. For instance, one can see that the non-dimensional stream-wise velocity reaches a steady value      
at $Wi = 1.5$; whereas, it becomes fluctuating in nature as the Weissenberg number is further increased to 2.5, sub Fig.~\ref{fig:velvstime}(c). These velocity fluctuations are governed by a two dominant frequencies (sub Fig.~\ref{fig:velvstime}(d)) as opposed to a range of frequency spectrum seen at $G = 0.28$ (sub Fig.~\ref{fig:velvstime}(b)) under otherwise identical conditions . Furthermore, the amplitude of these velocity fluctuations is seen to be very large in the latter case as compared to that seen in the former one. 

\begin{figure*}
    \centering
    \includegraphics[trim=0.5cm 0cm 6cm 0cm,clip,width=18cm]{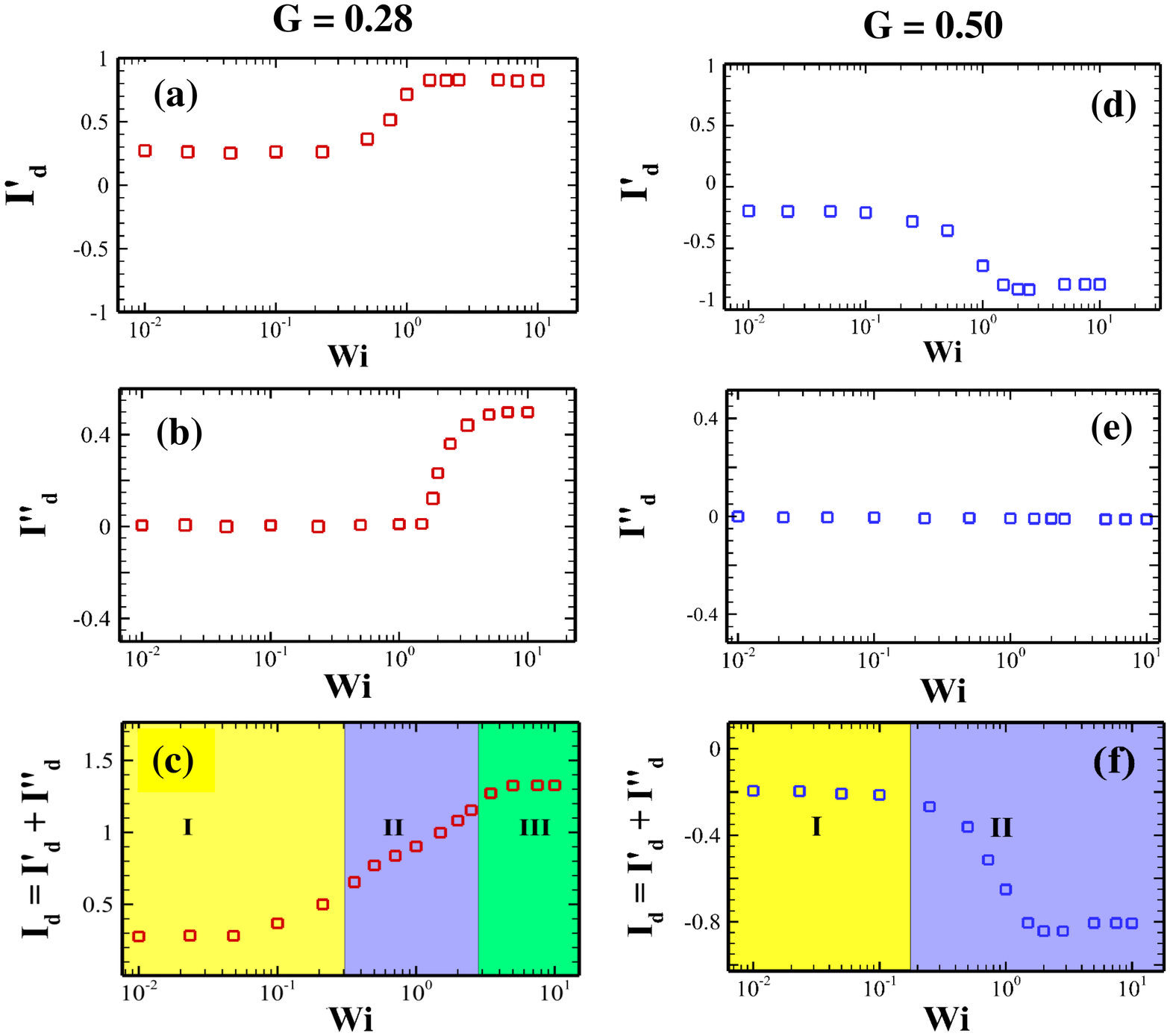}
    \caption{Variation of the flow asymmetry parameter for the two microcylinders case at $G = 0.28$ (a-c) and at $G = 0.5$ (d-f). In sub figure (c), (I) Newtonian like state (II) Diverging or 'D' state and (III) Asymmetric-diverging or 'AD' state, whereas in sub figure (f), (I) Newtonian like state and (II) converging or 'C' state.}
    \label{fig:IvsWidouble}
\end{figure*}
\begin{figure*}
    \centering
    \includegraphics[trim=0cm 8.5cm 0cm 0cm,clip,width=18cm]{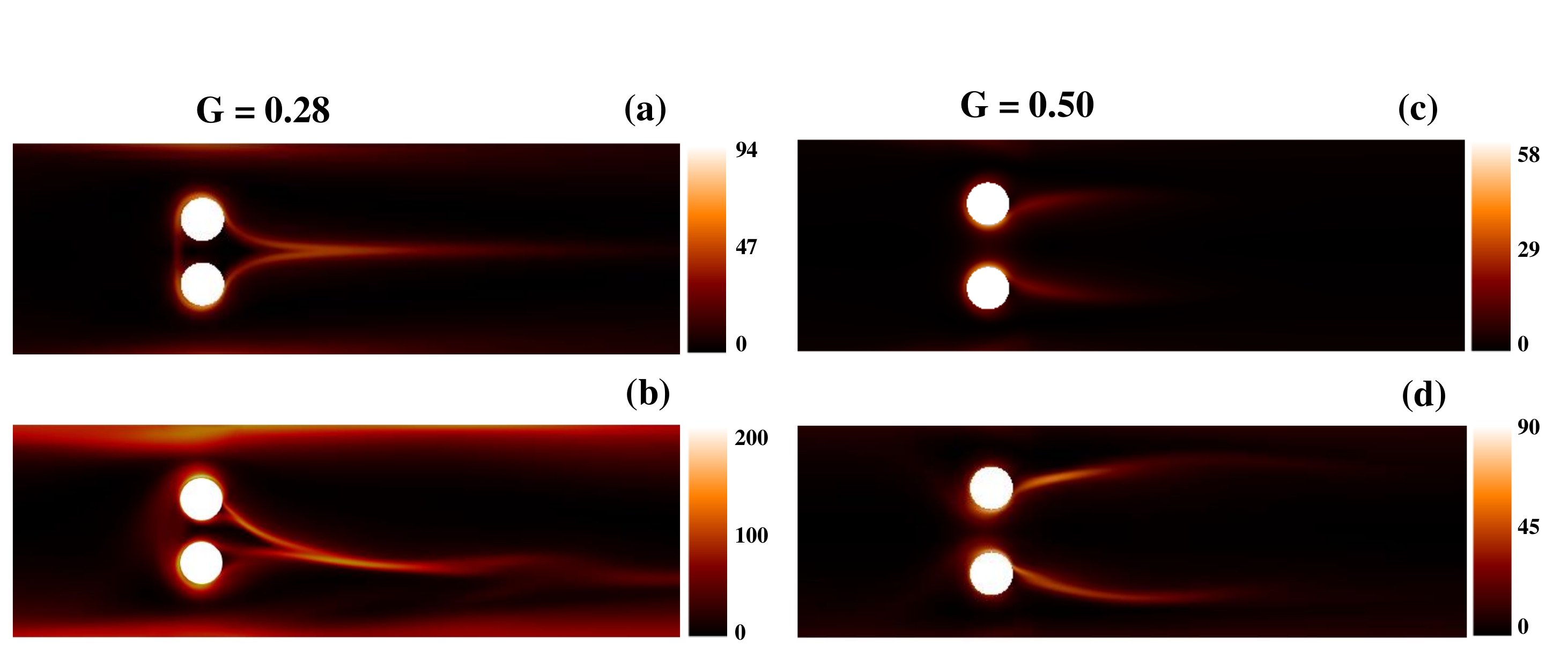}
    \caption{Variation of the principle stress difference for the two microcylinders case (a) G = 0.28, Wi = 1.0 (b) G = 0.28, Wi = 5.0 (c) G = 0.5, Wi = 1.0 (d) G = 0.5, Wi = 5.0.}
    \label{fig:PSDdouble}
\end{figure*}
Likewise Hopkins et al.~\cite{hopkins2020tristability}, we also calculate two asymmetrical parameters, namely, $I^{'}_{d}$ and $I^{''}_{d}$ to distinguish the flow states more quantitatively for the two microcylinders case. These are defined as follows:
\begin{equation}
    I^{'}_{d} = \frac{\frac{1}{2}\left( U_{X,u} + U_{X,l}\right) -U_{X,m}}{\frac{1}{2}\left( U_{X,u} + U_{X,l}\right) + U_{X,m}}
    \label{Id1}
\end{equation}
\begin{equation}
    I^{''}_{d} = \frac{ U_{X,u} - U_{X,l}}{ U_{X,u} + U_{X,l} + U_{X,m}}
    \label{Id1}
\end{equation}
In the above equations, $U_{X,u}$, $U_{X,l}$ and $U_{X,m}$ are the time-averaged stream-wise velocities obtained at the mid-points placed in the upper gap (between the top cylinder and channel wall), lower gap (between the bottom cylinder and lower channel wall) and in the gap in between the two cylinders, respectively. The variations of  $I^{'}_{d}$ and $I^{''}_{d}$ with the Weissenberg number are shown in sub Figs.~\ref{fig:IvsWidouble} (a-b) and (d-f) for the gap ratios of 0.28 and 0.5, respectively. The total asymmetry parameter $I_{d} = I^{'}_{d} + I^{''}_{d}$, showing the complete bifurcation diagram, is presented in sub Figs (c) and (f) at $G = 0.28$ and 0.50, respectively. The first transition in the value of $I^{'}_{d}$ occurs at  $Wi \approx 0.3$ when the flow transits from symmetric to diverging state (D). After this transition, as the Weissenberg number gradually increases, one can see that the value of $I^{'}_{d}$ also gradually increases, and ultimately leveling off to a value of 1, sub Fig.~\ref{fig:IvsWidouble}(a). This trend in $I^{'}_{d}$ thereby suggesting that almost no fluid passes in between the two cylinders as the Weissenberg number increases. The second transition in the flow state from the diverging (D) to asymmetric-diverging (AD) state occurs when the transition in the value of $I^{''}_{d}$ occurs at $Wi \approx 2.5$, sub Fig.~\ref{fig:IvsWidouble}(b). The complete bifurcation diagram at $G = 0.28$ is shown in sub Fig.~\ref{fig:IvsWidouble}(c) in terms of the variation of the total asymmetry parameter $I_{d}$ with $Wi$. It can be seen that the first bifurcation leads to $I_{d} \rightarrow 1 $, whereas the second bifurcation results in $I_{d} \rightarrow 1.5$. On the other hand, at $G = 0.50$, the first bifurcation occurs when the flow transits from symmetric to converging state (C) at $Wi \approx 0.15$, which can be marked by the transition of the value of $I^{'}_{d}$ in sub Fig.~\ref{fig:IvsWidouble}(d). As the Weissenberg number increases, the value of $I^{'}_{d}$ tends to -1, thereby suggesting that all of the fluid prefers to flow through the gap region in between the two cylinders. The value of $I^{''}_{d}$ almost remains zero over the whole range of the Weissenberg number considered (sub Fig.~\ref{fig:IvsWidouble}(e)), and hence, a second bifurcation is not observed at $G = 0.50$ as it was seen at $G = 0.28$. The complete bifurcation diagram for this gap ratio is depicted in sub Fig.~\ref{fig:IvsWidouble}(f).

To explain the formation of these different flow states in the case of flow past two microcylinders, the corresponding PSD plots at these two gap ratios are presented in Fig.~\ref{fig:PSDdouble}. At $G = 0.28$ and $Wi = 1.0$ at which 'D' states occurs, it can be observed that the gap in between the two cylinders is closed by a region of high PSD value (sub Fig.~\ref{fig:PSDdouble}(a)), thereby blocking the fluid to pass through this region. Furthermore, at this Weissenberg number, a long birefringent strand of high PSD value is also formed in the mid-horizontal plane downstream of the cylinders. As the Weissenberg number further increases to higher values, both the length and magnitude of this strand increase. A little and random lateral fluctuation in this strand in either $+Y$ or $-Y$ direction downstream of the cylinder blocks the flow of fluid in that direction, resulting in the formation of 'AD' state (sub Fig.~\ref{fig:PSDdouble}(b)). This is reminiscent of that seen in the case of single microcylinder. On the other hand, at $G = 0.5$, the velocity magnitude in between the two cylinders progressively increases as the Weissenberg number increases due to the shear-thinning property of the micellar solution, and hence more fluids prefer to pass through this area due to the formation of a low-viscosity region. As a result, the birefringent strands formed downstream of both the cylinders shift towards the channel walls (see sub Figs.~\ref{fig:PSDdouble}(c) and (d)), thereby blocking the fluid to pass through the gap regions in between the cylinder surface and channel wall. This facilitates more fluids to pass through the gap region in between the two cylinders. This effect gets accumulated as the Weissenberg number further increases, resulting in the formation of 'C' state. At this gap ratio, the space in between the two cylinders is not closed by a region of high PSD value (sub Fig.~\ref{fig:PSDdouble}(c)) as that seen at $G = 0.28$ which can block the flow, and therefore, the fluid can easily pass through this space. Likewise the single microcylinder case, we have again found that the flow bifurcation can be completely suppressed if the non-linear VCM model parameter $\xi$ increases to 0.1. In other words, if the micelles become progressively easier to break, this bifurcation in the two cylinders case can also be completely avoided due to the increase in the shear-thinning and decrease in the elastic effects, Fig.~\ref{fig:RheoFlow}. On the other hand, with a decreasing value of $\xi = 0.0001$ when the micelles become progressively harder to break, we have again observed the disappearance of these bifurcations in the flow irrespective of the gap ratio, due to an increase in the elastic and decrease in the shear-thinning effects, likewise we have seen for the single microcylinder case in the preceding subsection. 

\begin{figure*}
    \centering
    \includegraphics[trim=4cm 8.5cm 0.5cm 6cm,clip,width=12cm]{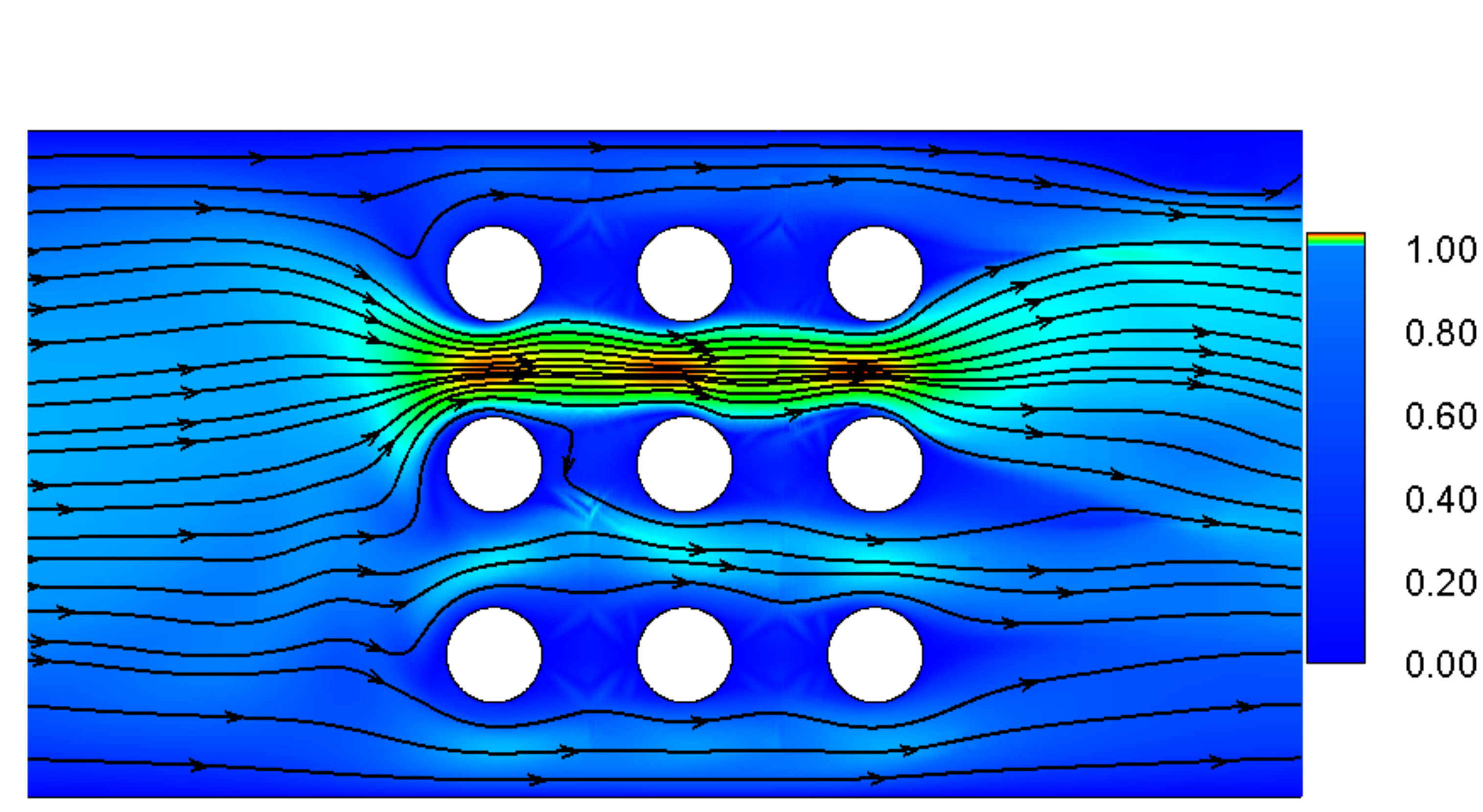}
    \caption{Streamline and velocity magnitude plots for the flow of WLM solutions through an ordered porous structure consisting of a microchannel with multiple microcylinders placed in it at $Wi = 4$ and $\xi = 0.01$.}
    \label{fig:FinalFigure}
\end{figure*}
All these results presented and discussed here for single and two microcylinders cases now can facilitate the understanding of the selection of a preferential path or lane of a viscoelastic fluid during its flow through either an ordered or disordered porous matrix observed in many prior experiments~\cite{de2017lane,de2018flow,walkama2020disorder,eberhard2020mapping,muller1998optical}. The onset of this phenomena happens due to the occurrence of the flow bifurcation (either 'A' or 'AD' or 'C' state) resulting from the interaction between the shear-thinning properties of the micellar solution and elastic stresses generated in the system, as explained above. Once the fluid prefers to flow through a particular gap region in the porous media due to the flow bifurcation, it then forms a lane or path as moves forward. To demonstrate this, we have carried out further numerical simulations for an ordered porous matrix created by placing nine microcylinders in a microchannel, as schematically shown in Fig.~\ref{fig:FinalFigure}. One can clearly see the formation of a preferential path or lane during the flow of micellar solutions through this ordered porous matrix.      

\section{\label{Con} Conclusions}
In this study, the flow phenomena of wormlike micellar solutions (WLM) past a single and two vertically aligned microcylinders placed in a rectangular channel is numerically investigated in detail in the creeping flow regime. The two-species Vasquez-Cook-McKinley (VCM) constitutive model, which includes both the breakage and reformation dynamics of micelles, is used to characterize the rheological behaviour of WLM solutions. At low Weissenberg numbers, the flow dynamics is found to be steady and symmetric for both the single and two microcylinders cases regardless of the blockage ($BR = \frac{D}{H}$, where $D$ is the cylinder diameter and $H$ is the channel height) and gap ($G = \frac{S_{1}}{S_{1} + S_{2}}$ where $S_{1}$ is the distance between the two cylinder and $S_{2}$ is the distance between the channel wall and cylinder surface) ratio, likewise seen for simple Newtonian fluids in the creeping flow regime. However, as the Weissenberg number gradually increases to high values, the flow features become rich in physics and also become dependent on the blockage and gap ratio. For instance, in the case of a single microcylinder, a range of blockage ratio is found at which an asymmetric flow is seen to exist due to the occurrence of a supercritical pitchfork bifurcation in the flow field. At higher blockage ratios, a resymmetrization in the flow field happens. Along with this, a transition for a wide range of flow states is found as the blockage ratio gradually increases. However, all these observations are found to be a strong function of the non-linear VCM model parameter $\xi$ which basically indicates how easy or hard to break a micelle. As the value of $\xi$ increases or it becomes progressively easier to break a micelle (thereby increasing the shear-thinning tendency and decreasing the elastic property), the asymmetric flow is totally disappeared irrespective of the blockage ratio. On the other hand, as the micelles become progressively hard to break or decreasing value of $\xi$, the asymmetric flow again disappears. This suggests that there is a range of value of $\xi$ present at which both the shear-thinning properties of the micellar solutions and an accumulation of the elastic stresses downstream of the cylinder become significant, which thereby resulting in an asymmetric flow in the system. This observation is in line with that presented earlier for the flow of either WLM~\cite{haward2019flow} or polymer~\cite{haward2020asymmetric} solutions past a cylinder.

In the case of two microcylinders aligned vertically to each other, once again, the flow field of WLM solutions seems like Newtonian fluids, i.e., steady and symmetric at low Weissenberg numbers. As it gradually increases to higher values, three distinct flow states are observed in the system, namely, diverging ('D') state at which most of the fluids pass through the gaps in between the cylinder surface and channel walls, asymmetric-diverging ('AD') state at which the micellar solution prefers to flow through the gap in between either the top channel wall and cylinder surface or the bottom channel wall and cylinder surface, and converging state ('C') at which most of the fluids flow through the gap in between the two cylinders. All these flow states are also observed in recent experiments~\cite{hopkins2020tristability} in the case of two microcylinders dealing with WLM solutions. We have found that the occurrence of any of these states is strongly dependent upon the values of the gap ratio and non-linear VCM model parameter $\xi$. Once again, the reason behind the formation of these flow states lies to the fact of the interaction between the shear-thinning properties and accumulation of the elastic stresses downstream of the cylinders. Therefore, the formation of any of these flow states can be controlled by changing the scission energy needed to break a micelle or the value of $\xi$. We have found the occurrence of a bistable state at $G = 0.28$ and a single stable state at $G = 0.50$. In between these two $G$ values, one can expect a critical gap ratio at which all these three states (tristable) co-exist together as seen in the recent experiments~\cite{hopkins2020tristability}. However, we are unable to find out that critical value of the gap ratio in the present simulations. 

Finally, based on the results and explanations presented herein for the single and two microcylinders, we have provided the reason behind the formation of preferential paths or lanes during the flow of either WLM or polymer solutions through a porous media, as observed in many earlier experiments~\cite{de2018flow,de2017lane,walkama2020disorder,eberhard2020mapping}. The onset of this phenomena happens due to the occurrence of the flow bifurcation (either 'D' or 'AD' or 'C' state) resulting from the interaction between the shear-thinning properties of the viscoelastic fluid and elastic stresses generated in the system. This lane formation can happen in both polymer and wormlike micellar solutions as long as the solution exhibits both the shear-thinning properties and accumulates sufficient elastic stresses downstream of the obstacle, as it was experimentally observed in both the solutions. For a wormlike micellar solution, both these shear-thinning and elastic properties are influenced by the fact that how easy or hard to break a micelle (by the non-linear parameter $\xi$ in the case of VCM model), and hence, one can say that the lane formation in wormlike micellar solution is indirectly dependent on the breakage and reformation dynamics of micelles.  

\section{Acknowledgements}
The authors would like to thank IIT Ropar for providing the funding through the ISIRD research grant (Establishment 1/2018/IITRPR/921) to carry out this work. 

\section{Availability of data}
The data that supports the findings of this study are available within the article.
\nocite{*}
\bibliography{aipsamp}

\providecommand{\noopsort}[1]{}\providecommand{\singleletter}[1]{#1}%
\begin{thebibliography}{50}%
\makeatletter
\providecommand \@ifxundefined [1]{%
 \@ifx{#1\undefined}
}%
\providecommand \@ifnum [1]{%
 \ifnum #1\expandafter \@firstoftwo
 \else \expandafter \@secondoftwo
 \fi
}%
\providecommand \@ifx [1]{%
 \ifx #1\expandafter \@firstoftwo
 \else \expandafter \@secondoftwo
 \fi
}%
\providecommand \natexlab [1]{#1}%
\providecommand \enquote  [1]{``#1''}%
\providecommand \bibnamefont  [1]{#1}%
\providecommand \bibfnamefont [1]{#1}%
\providecommand \citenamefont [1]{#1}%
\providecommand \href@noop [0]{\@secondoftwo}%
\providecommand \href [0]{\begingroup \@sanitize@url \@href}%
\providecommand \@href[1]{\@@startlink{#1}\@@href}%
\providecommand \@@href[1]{\endgroup#1\@@endlink}%
\providecommand \@sanitize@url [0]{\catcode `\\12\catcode `\$12\catcode
  `\&12\catcode `\#12\catcode `\^12\catcode `\_12\catcode `\%12\relax}%
\providecommand \@@startlink[1]{}%
\providecommand \@@endlink[0]{}%
\providecommand \url  [0]{\begingroup\@sanitize@url \@url }%
\providecommand \@url [1]{\endgroup\@href {#1}{\urlprefix }}%
\providecommand \urlprefix  [0]{URL }%
\providecommand \Eprint [0]{\href }%
\providecommand \doibase [0]{http://dx.doi.org/}%
\providecommand \selectlanguage [0]{\@gobble}%
\providecommand \bibinfo  [0]{\@secondoftwo}%
\providecommand \bibfield  [0]{\@secondoftwo}%
\providecommand \translation [1]{[#1]}%
\providecommand \BibitemOpen [0]{}%
\providecommand \bibitemStop [0]{}%
\providecommand \bibitemNoStop [0]{.\EOS\space}%
\providecommand \EOS [0]{\spacefactor3000\relax}%
\providecommand \BibitemShut  [1]{\csname bibitem#1\endcsname}%
\let\auto@bib@innerbib\@empty
\bibitem [{\citenamefont {Dreiss}(2007)}]{dreiss2007wormlike}%
  \BibitemOpen
  \bibfield  {author} {\bibinfo {author} {\bibfnamefont {C.~A.}\ \bibnamefont
  {Dreiss}},\ }\bibfield  {title} {\enquote {\bibinfo {title} {Wormlike
  micelles: where do we stand? recent developments, linear rheology and
  scattering techniques},}\ }\href@noop {} {\bibfield  {journal} {\bibinfo
  {journal} {Soft Matt.}\ }\textbf {\bibinfo {volume} {3}},\ \bibinfo {pages}
  {956--970} (\bibinfo {year} {2007})}\BibitemShut {NoStop}%
\bibitem [{\citenamefont {Dreiss}\ and\ \citenamefont
  {Feng}(2017)}]{dreiss2017wormlike}%
  \BibitemOpen
  \bibfield  {author} {\bibinfo {author} {\bibfnamefont {C.~A.}\ \bibnamefont
  {Dreiss}}\ and\ \bibinfo {author} {\bibfnamefont {Y.}~\bibnamefont {Feng}},\
  }\href@noop {} {\emph {\bibinfo {title} {Wormlike {M}icelles: {A}dvances in
  {S}ystems, {C}haracterisation and {A}pplications}}},\ Vol.~\bibinfo {volume}
  {6}\ (\bibinfo  {publisher} {Royal Society of Chemistry},\ \bibinfo {year}
  {2017})\BibitemShut {NoStop}%
\bibitem [{\citenamefont {Yang}(2002)}]{yang2002viscoelastic}%
  \BibitemOpen
  \bibfield  {author} {\bibinfo {author} {\bibfnamefont {J.}~\bibnamefont
  {Yang}},\ }\bibfield  {title} {\enquote {\bibinfo {title} {Viscoelastic
  wormlike micelles and their applications},}\ }\href@noop {} {\bibfield
  {journal} {\bibinfo  {journal} {Cur. Opi. Col. Int. Sci.}\ }\textbf {\bibinfo
  {volume} {7}},\ \bibinfo {pages} {276--281} (\bibinfo {year}
  {2002})}\BibitemShut {NoStop}%
\bibitem [{\citenamefont {Walker}(2001)}]{walker2001rheology}%
  \BibitemOpen
  \bibfield  {author} {\bibinfo {author} {\bibfnamefont {L.~M.}\ \bibnamefont
  {Walker}},\ }\bibfield  {title} {\enquote {\bibinfo {title} {Rheology and
  structure of worm-like micelles},}\ }\href@noop {} {\bibfield  {journal}
  {\bibinfo  {journal} {Cur. Opi. Col. Int. Sci.}\ }\textbf {\bibinfo {volume}
  {6}},\ \bibinfo {pages} {451--456} (\bibinfo {year} {2001})}\BibitemShut
  {NoStop}%
\bibitem [{\citenamefont {Rothstein}(2008)}]{rothstein2008strong}%
  \BibitemOpen
  \bibfield  {author} {\bibinfo {author} {\bibfnamefont {J.~P.}\ \bibnamefont
  {Rothstein}},\ }\bibfield  {title} {\enquote {\bibinfo {title} {Strong flows
  of viscoelastic wormlike micelle solutions},}\ }\href@noop {} {\bibfield
  {journal} {\bibinfo  {journal} {Rheol. Rev}\ }\textbf {\bibinfo {volume}
  {2008}},\ \bibinfo {pages} {1--46} (\bibinfo {year} {2008})}\BibitemShut
  {NoStop}%
\bibitem [{\citenamefont {Rothstein}(2003)}]{rothstein2003transient}%
  \BibitemOpen
  \bibfield  {author} {\bibinfo {author} {\bibfnamefont {J.~P.}\ \bibnamefont
  {Rothstein}},\ }\bibfield  {title} {\enquote {\bibinfo {title} {Transient
  extensional rheology of wormlike micelle solutions},}\ }\href@noop {}
  {\bibfield  {journal} {\bibinfo  {journal} {J. Rheol.}\ }\textbf {\bibinfo
  {volume} {47}},\ \bibinfo {pages} {1227--1247} (\bibinfo {year}
  {2003})}\BibitemShut {NoStop}%
\bibitem [{\citenamefont {Berret}(1997)}]{berret1997transient}%
  \BibitemOpen
  \bibfield  {author} {\bibinfo {author} {\bibfnamefont {J.-F.}\ \bibnamefont
  {Berret}},\ }\bibfield  {title} {\enquote {\bibinfo {title} {Transient
  rheology of wormlike micelles},}\ }\href@noop {} {\bibfield  {journal}
  {\bibinfo  {journal} {Langmuir}\ }\textbf {\bibinfo {volume} {13}},\ \bibinfo
  {pages} {2227--2234} (\bibinfo {year} {1997})}\BibitemShut {NoStop}%
\bibitem [{\citenamefont {Schramm}(2000)}]{schramm2000surfactants}%
  \BibitemOpen
  \bibfield  {author} {\bibinfo {author} {\bibfnamefont {L.~L.}\ \bibnamefont
  {Schramm}},\ }\href@noop {} {\emph {\bibinfo {title} {Surfactants:
  {F}undamentals and {A}pplications in the {P}etroleum {I}ndustry}}}\ (\bibinfo
   {publisher} {Cambridge University Press},\ \bibinfo {year}
  {2000})\BibitemShut {NoStop}%
\bibitem [{\citenamefont {M{\"o}bius}, \citenamefont {Miller},\ and\
  \citenamefont {Fainerman}(2001)}]{mobius2001surfactants}%
  \BibitemOpen
  \bibfield  {author} {\bibinfo {author} {\bibfnamefont {D.}~\bibnamefont
  {M{\"o}bius}}, \bibinfo {author} {\bibfnamefont {R.}~\bibnamefont {Miller}},
  \ and\ \bibinfo {author} {\bibfnamefont {V.~B.}\ \bibnamefont {Fainerman}},\
  }\href@noop {} {\emph {\bibinfo {title} {Surfactants: {C}hemistry,
  {I}nterfacial {P}roperties, {A}pplications}}}\ (\bibinfo  {publisher}
  {Elsevier},\ \bibinfo {year} {2001})\BibitemShut {NoStop}%
\bibitem [{\citenamefont {Raffa}\ \emph {et~al.}(2015)\citenamefont {Raffa},
  \citenamefont {Wever}, \citenamefont {Picchioni},\ and\ \citenamefont
  {Broekhuis}}]{raffa2015polymeric}%
  \BibitemOpen
  \bibfield  {author} {\bibinfo {author} {\bibfnamefont {P.}~\bibnamefont
  {Raffa}}, \bibinfo {author} {\bibfnamefont {D.~A.~Z.}\ \bibnamefont {Wever}},
  \bibinfo {author} {\bibfnamefont {F.}~\bibnamefont {Picchioni}}, \ and\
  \bibinfo {author} {\bibfnamefont {A.~A.}\ \bibnamefont {Broekhuis}},\
  }\bibfield  {title} {\enquote {\bibinfo {title} {Polymeric surfactants:
  synthesis, properties, and links to applications},}\ }\href@noop {}
  {\bibfield  {journal} {\bibinfo  {journal} {Chem. Reviews}\ }\textbf
  {\bibinfo {volume} {115}},\ \bibinfo {pages} {8504--8563} (\bibinfo {year}
  {2015})}\BibitemShut {NoStop}%
\bibitem [{\citenamefont {De}\ \emph {et~al.}(2018)\citenamefont {De},
  \citenamefont {Koesen}, \citenamefont {Maitri}, \citenamefont {Golombok},
  \citenamefont {Padding},\ and\ \citenamefont {van Santvoort}}]{de2018flow}%
  \BibitemOpen
  \bibfield  {author} {\bibinfo {author} {\bibfnamefont {S.}~\bibnamefont
  {De}}, \bibinfo {author} {\bibfnamefont {S.~P.}\ \bibnamefont {Koesen}},
  \bibinfo {author} {\bibfnamefont {R.~V.}\ \bibnamefont {Maitri}}, \bibinfo
  {author} {\bibfnamefont {M.}~\bibnamefont {Golombok}}, \bibinfo {author}
  {\bibfnamefont {J.~T.}\ \bibnamefont {Padding}}, \ and\ \bibinfo {author}
  {\bibfnamefont {J.~F.~M.}\ \bibnamefont {van Santvoort}},\ }\bibfield
  {title} {\enquote {\bibinfo {title} {Flow of viscoelastic surfactants through
  porous media},}\ }\href@noop {} {\bibfield  {journal} {\bibinfo  {journal}
  {AIChE J.}\ }\textbf {\bibinfo {volume} {64}},\ \bibinfo {pages} {773--781}
  (\bibinfo {year} {2018})}\BibitemShut {NoStop}%
\bibitem [{\citenamefont {De}\ \emph {et~al.}(2017)\citenamefont {De},
  \citenamefont {Van Der~Schaaf}, \citenamefont {Deen}, \citenamefont
  {Kuipers}, \citenamefont {Peters},\ and\ \citenamefont
  {Padding}}]{de2017lane}%
  \BibitemOpen
  \bibfield  {author} {\bibinfo {author} {\bibfnamefont {S.}~\bibnamefont
  {De}}, \bibinfo {author} {\bibfnamefont {J.}~\bibnamefont {Van Der~Schaaf}},
  \bibinfo {author} {\bibfnamefont {N.~G.}\ \bibnamefont {Deen}}, \bibinfo
  {author} {\bibfnamefont {J.~A.~M.}\ \bibnamefont {Kuipers}}, \bibinfo
  {author} {\bibfnamefont {E.~A. J.~F.}\ \bibnamefont {Peters}}, \ and\
  \bibinfo {author} {\bibfnamefont {J.~T.}\ \bibnamefont {Padding}},\
  }\bibfield  {title} {\enquote {\bibinfo {title} {Lane change in flows through
  pillared microchannels},}\ }\href@noop {} {\bibfield  {journal} {\bibinfo
  {journal} {Phys. Fluids}\ }\textbf {\bibinfo {volume} {29}},\ \bibinfo
  {pages} {113102} (\bibinfo {year} {2017})}\BibitemShut {NoStop}%
\bibitem [{\citenamefont {M{\"u}ller}, \citenamefont {Vorwerk},\ and\
  \citenamefont {Brunn}(1998)}]{muller1998optical}%
  \BibitemOpen
  \bibfield  {author} {\bibinfo {author} {\bibfnamefont {M.}~\bibnamefont
  {M{\"u}ller}}, \bibinfo {author} {\bibfnamefont {J.}~\bibnamefont {Vorwerk}},
  \ and\ \bibinfo {author} {\bibfnamefont {P.}~\bibnamefont {Brunn}},\
  }\bibfield  {title} {\enquote {\bibinfo {title} {Optical studies of local
  flow behaviour of a non-newtonian fluid inside a porous medium},}\
  }\href@noop {} {\bibfield  {journal} {\bibinfo  {journal} {Rheol. Acta}\
  }\textbf {\bibinfo {volume} {37}},\ \bibinfo {pages} {189--194} (\bibinfo
  {year} {1998})}\BibitemShut {NoStop}%
\bibitem [{\citenamefont {Walkama}, \citenamefont {Waisbord},\ and\
  \citenamefont {Guasto}(2020)}]{walkama2020disorder}%
  \BibitemOpen
  \bibfield  {author} {\bibinfo {author} {\bibfnamefont {D.~M.}\ \bibnamefont
  {Walkama}}, \bibinfo {author} {\bibfnamefont {N.}~\bibnamefont {Waisbord}}, \
  and\ \bibinfo {author} {\bibfnamefont {J.~S.}\ \bibnamefont {Guasto}},\
  }\bibfield  {title} {\enquote {\bibinfo {title} {Disorder suppresses chaos in
  viscoelastic flows},}\ }\href@noop {} {\bibfield  {journal} {\bibinfo
  {journal} {Phys. Rev. Lett.}\ }\textbf {\bibinfo {volume} {124}},\ \bibinfo
  {pages} {164501} (\bibinfo {year} {2020})}\BibitemShut {NoStop}%
\bibitem [{\citenamefont {Eberhard}\ \emph {et~al.}(2020)\citenamefont
  {Eberhard}, \citenamefont {Seybold}, \citenamefont {Secchi}, \citenamefont
  {Jim{\'e}nez-Mart{\'\i}nez}, \citenamefont {R{\"u}hs}, \citenamefont {Ofner},
  \citenamefont {Andrade},\ and\ \citenamefont
  {Holzner}}]{eberhard2020mapping}%
  \BibitemOpen
  \bibfield  {author} {\bibinfo {author} {\bibfnamefont {U.}~\bibnamefont
  {Eberhard}}, \bibinfo {author} {\bibfnamefont {H.}~\bibnamefont {Seybold}},
  \bibinfo {author} {\bibfnamefont {E.}~\bibnamefont {Secchi}}, \bibinfo
  {author} {\bibfnamefont {J.}~\bibnamefont {Jim{\'e}nez-Mart{\'\i}nez}},
  \bibinfo {author} {\bibfnamefont {P.}~\bibnamefont {R{\"u}hs}}, \bibinfo
  {author} {\bibfnamefont {A.}~\bibnamefont {Ofner}}, \bibinfo {author}
  {\bibfnamefont {J.}~\bibnamefont {Andrade}}, \ and\ \bibinfo {author}
  {\bibfnamefont {M.}~\bibnamefont {Holzner}},\ }\bibfield  {title} {\enquote
  {\bibinfo {title} {Mapping the local viscosity of non-{N}ewtonian fluids
  flowing through disordered porous structures},}\ }\href@noop {} {\bibfield
  {journal} {\bibinfo  {journal} {Sci. Reports}\ }\textbf {\bibinfo {volume}
  {10}},\ \bibinfo {pages} {1--12} (\bibinfo {year} {2020})}\BibitemShut
  {NoStop}%
\bibitem [{\citenamefont {Alves}, \citenamefont {Pinho},\ and\ \citenamefont
  {Oliveira}(2001)}]{alves2001flow}%
  \BibitemOpen
  \bibfield  {author} {\bibinfo {author} {\bibfnamefont {M.~A.}\ \bibnamefont
  {Alves}}, \bibinfo {author} {\bibfnamefont {F.~T.}\ \bibnamefont {Pinho}}, \
  and\ \bibinfo {author} {\bibfnamefont {P.~J.}\ \bibnamefont {Oliveira}},\
  }\bibfield  {title} {\enquote {\bibinfo {title} {The flow of viscoelastic
  fluids past a cylinder: finite-volume high-resolution methods},}\ }\href@noop
  {} {\bibfield  {journal} {\bibinfo  {journal} {J. Non-Newt. Fluid Mech.}\
  }\textbf {\bibinfo {volume} {97}},\ \bibinfo {pages} {207--232} (\bibinfo
  {year} {2001})}\BibitemShut {NoStop}%
\bibitem [{\citenamefont {McKinley}, \citenamefont {Armstrong},\ and\
  \citenamefont {Brown}(1993)}]{mckinley1993wake}%
  \BibitemOpen
  \bibfield  {author} {\bibinfo {author} {\bibfnamefont {G.~H.}\ \bibnamefont
  {McKinley}}, \bibinfo {author} {\bibfnamefont {R.~C.}\ \bibnamefont
  {Armstrong}}, \ and\ \bibinfo {author} {\bibfnamefont {R.}~\bibnamefont
  {Brown}},\ }\bibfield  {title} {\enquote {\bibinfo {title} {The wake
  instability in viscoelastic flow past confined circular cylinders},}\
  }\href@noop {} {\bibfield  {journal} {\bibinfo  {journal} {Phil. Tran. Royal
  Society of London. Series A: Phys. Eng. Sci.}\ }\textbf {\bibinfo {volume}
  {344}},\ \bibinfo {pages} {265--304} (\bibinfo {year} {1993})}\BibitemShut
  {NoStop}%
\bibitem [{\citenamefont {Hu}\ and\ \citenamefont
  {Joseph}(1990)}]{hu1990numerical}%
  \BibitemOpen
  \bibfield  {author} {\bibinfo {author} {\bibfnamefont {H.~H.}\ \bibnamefont
  {Hu}}\ and\ \bibinfo {author} {\bibfnamefont {D.~D.}\ \bibnamefont
  {Joseph}},\ }\bibfield  {title} {\enquote {\bibinfo {title} {Numerical
  simulation of viscoelastic flow past a cylinder},}\ }\href@noop {} {\bibfield
   {journal} {\bibinfo  {journal} {J. Non-Newt. Fluid Mech.}\ }\textbf
  {\bibinfo {volume} {37}},\ \bibinfo {pages} {347--377} (\bibinfo {year}
  {1990})}\BibitemShut {NoStop}%
\bibitem [{\citenamefont {Shiang}\ \emph {et~al.}(1997)\citenamefont {Shiang},
  \citenamefont {Lin}, \citenamefont {{\"O}ztekin},\ and\ \citenamefont
  {Rockwell}}]{shiang1997viscoelastic}%
  \BibitemOpen
  \bibfield  {author} {\bibinfo {author} {\bibfnamefont {A.~H.}\ \bibnamefont
  {Shiang}}, \bibinfo {author} {\bibfnamefont {J.~C.}\ \bibnamefont {Lin}},
  \bibinfo {author} {\bibfnamefont {A.}~\bibnamefont {{\"O}ztekin}}, \ and\
  \bibinfo {author} {\bibfnamefont {D.}~\bibnamefont {Rockwell}},\ }\bibfield
  {title} {\enquote {\bibinfo {title} {Viscoelastic flow around a confined
  circular cylinder: measurements using high-image-density particle image
  velocimetry},}\ }\href@noop {} {\bibfield  {journal} {\bibinfo  {journal} {J.
  Non-Newt. Fluid Mech.}\ }\textbf {\bibinfo {volume} {73}},\ \bibinfo {pages}
  {29--49} (\bibinfo {year} {1997})}\BibitemShut {NoStop}%
\bibitem [{\citenamefont {Qin}\ \emph {et~al.}(2019)\citenamefont {Qin},
  \citenamefont {Salipante}, \citenamefont {Hudson},\ and\ \citenamefont
  {Arratia}}]{qin2019upstream}%
  \BibitemOpen
  \bibfield  {author} {\bibinfo {author} {\bibfnamefont {B.}~\bibnamefont
  {Qin}}, \bibinfo {author} {\bibfnamefont {P.~F.}\ \bibnamefont {Salipante}},
  \bibinfo {author} {\bibfnamefont {S.~D.}\ \bibnamefont {Hudson}}, \ and\
  \bibinfo {author} {\bibfnamefont {P.~E.}\ \bibnamefont {Arratia}},\
  }\bibfield  {title} {\enquote {\bibinfo {title} {Upstream vortex and elastic
  wave in the viscoelastic flow around a confined cylinder},}\ }\href@noop {}
  {\bibfield  {journal} {\bibinfo  {journal} {J. Fluid Mech.}\ }\textbf
  {\bibinfo {volume} {864}} (\bibinfo {year} {2019})}\BibitemShut {NoStop}%
\bibitem [{\citenamefont {Moss}\ and\ \citenamefont
  {Rothstein}(2010)}]{moss2010flow}%
  \BibitemOpen
  \bibfield  {author} {\bibinfo {author} {\bibfnamefont {G.~R.}\ \bibnamefont
  {Moss}}\ and\ \bibinfo {author} {\bibfnamefont {J.~P.}\ \bibnamefont
  {Rothstein}},\ }\bibfield  {title} {\enquote {\bibinfo {title} {Flow of
  wormlike micelle solutions past a confined circular cylinder},}\ }\href@noop
  {} {\bibfield  {journal} {\bibinfo  {journal} {J. Non-Newt. Fluid Mech.}\
  }\textbf {\bibinfo {volume} {165}},\ \bibinfo {pages} {1505--1515} (\bibinfo
  {year} {2010})}\BibitemShut {NoStop}%
\bibitem [{\citenamefont {Zhao}, \citenamefont {Shen},\ and\ \citenamefont
  {Haward}(2016)}]{zhao2016flow}%
  \BibitemOpen
  \bibfield  {author} {\bibinfo {author} {\bibfnamefont {Y.}~\bibnamefont
  {Zhao}}, \bibinfo {author} {\bibfnamefont {A.~Q.}\ \bibnamefont {Shen}}, \
  and\ \bibinfo {author} {\bibfnamefont {S.~J.}\ \bibnamefont {Haward}},\
  }\bibfield  {title} {\enquote {\bibinfo {title} {Flow of wormlike micellar
  solutions around confined microfluidic cylinders},}\ }\href@noop {}
  {\bibfield  {journal} {\bibinfo  {journal} {Soft Matt.}\ }\textbf {\bibinfo
  {volume} {12}},\ \bibinfo {pages} {8666--8681} (\bibinfo {year}
  {2016})}\BibitemShut {NoStop}%
\bibitem [{\citenamefont {Haward}\ \emph {et~al.}(2019)\citenamefont {Haward},
  \citenamefont {Kitajima}, \citenamefont {Toda-Peters}, \citenamefont
  {Takahashi},\ and\ \citenamefont {Shen}}]{haward2019flow}%
  \BibitemOpen
  \bibfield  {author} {\bibinfo {author} {\bibfnamefont {S.~J.}\ \bibnamefont
  {Haward}}, \bibinfo {author} {\bibfnamefont {N.}~\bibnamefont {Kitajima}},
  \bibinfo {author} {\bibfnamefont {K.}~\bibnamefont {Toda-Peters}}, \bibinfo
  {author} {\bibfnamefont {T.}~\bibnamefont {Takahashi}}, \ and\ \bibinfo
  {author} {\bibfnamefont {A.~Q.}\ \bibnamefont {Shen}},\ }\bibfield  {title}
  {\enquote {\bibinfo {title} {Flow of wormlike micellar solutions around
  microfluidic cylinders with high aspect ratio and low blockage ratio},}\
  }\href@noop {} {\bibfield  {journal} {\bibinfo  {journal} {Soft Matt.}\
  }\textbf {\bibinfo {volume} {15}},\ \bibinfo {pages} {1927--1941} (\bibinfo
  {year} {2019})}\BibitemShut {NoStop}%
\bibitem [{\citenamefont {Khan}\ and\ \citenamefont
  {Sasmal}(2020)}]{khan2020effect}%
  \BibitemOpen
  \bibfield  {author} {\bibinfo {author} {\bibfnamefont {M.~B.}\ \bibnamefont
  {Khan}}\ and\ \bibinfo {author} {\bibfnamefont {C.}~\bibnamefont {Sasmal}},\
  }\bibfield  {title} {\enquote {\bibinfo {title} {Effect of chain scission on
  flow characteristics of wormlike micellar solutions past a confined
  microfluidic cylinder: A numerical analysis},}\ }\href@noop {} {\bibfield
  {journal} {\bibinfo  {journal} {Soft Matt.}\ }\textbf {\bibinfo {volume}
  {16}},\ \bibinfo {pages} {5261--5272} (\bibinfo {year} {2020})}\BibitemShut
  {NoStop}%
\bibitem [{\citenamefont {Varchanis}\ \emph {et~al.}(2020)\citenamefont
  {Varchanis}, \citenamefont {Hopkins}, \citenamefont {Shen}, \citenamefont
  {Tsamopoulos},\ and\ \citenamefont {Haward}}]{varchanis2020asymmetric}%
  \BibitemOpen
  \bibfield  {author} {\bibinfo {author} {\bibfnamefont {S.}~\bibnamefont
  {Varchanis}}, \bibinfo {author} {\bibfnamefont {C.~C.}\ \bibnamefont
  {Hopkins}}, \bibinfo {author} {\bibfnamefont {A.~Q.}\ \bibnamefont {Shen}},
  \bibinfo {author} {\bibfnamefont {J.}~\bibnamefont {Tsamopoulos}}, \ and\
  \bibinfo {author} {\bibfnamefont {S.~J.}\ \bibnamefont {Haward}},\ }\bibfield
   {title} {\enquote {\bibinfo {title} {Asymmetric flows of complex fluids past
  confined cylinders: {A} comprehensive numerical study with experimental
  validation},}\ }\href@noop {} {\bibfield  {journal} {\bibinfo  {journal}
  {Phys. Fluids}\ }\textbf {\bibinfo {volume} {32}},\ \bibinfo {pages} {053103}
  (\bibinfo {year} {2020})}\BibitemShut {NoStop}%
\bibitem [{\citenamefont {Haward}, \citenamefont {Toda-Peters},\ and\
  \citenamefont {Shen}(2018)}]{haward2018steady}%
  \BibitemOpen
  \bibfield  {author} {\bibinfo {author} {\bibfnamefont {S.~J.}\ \bibnamefont
  {Haward}}, \bibinfo {author} {\bibfnamefont {K.}~\bibnamefont {Toda-Peters}},
  \ and\ \bibinfo {author} {\bibfnamefont {A.~Q.}\ \bibnamefont {Shen}},\
  }\bibfield  {title} {\enquote {\bibinfo {title} {Steady viscoelastic flow
  around high-aspect-ratio, low-blockage-ratio microfluidic cylinders},}\
  }\href@noop {} {\bibfield  {journal} {\bibinfo  {journal} {J. Non-Newt. Fluid
  Mech.}\ }\textbf {\bibinfo {volume} {254}},\ \bibinfo {pages} {23--35}
  (\bibinfo {year} {2018})}\BibitemShut {NoStop}%
\bibitem [{\citenamefont {Varshney}\ and\ \citenamefont
  {Steinberg}(2017)}]{varshney2017elastic}%
  \BibitemOpen
  \bibfield  {author} {\bibinfo {author} {\bibfnamefont {A.}~\bibnamefont
  {Varshney}}\ and\ \bibinfo {author} {\bibfnamefont {V.}~\bibnamefont
  {Steinberg}},\ }\bibfield  {title} {\enquote {\bibinfo {title} {Elastic wake
  instabilities in a creeping flow between two obstacles},}\ }\href@noop {}
  {\bibfield  {journal} {\bibinfo  {journal} {Phys. Rev. Fluids}\ }\textbf
  {\bibinfo {volume} {2}},\ \bibinfo {pages} {051301} (\bibinfo {year}
  {2017})}\BibitemShut {NoStop}%
\bibitem [{\citenamefont {Cressman}, \citenamefont {Bailey},\ and\
  \citenamefont {Goldburg}(2001)}]{cressman2001modification}%
  \BibitemOpen
  \bibfield  {author} {\bibinfo {author} {\bibfnamefont {J.~R.}\ \bibnamefont
  {Cressman}}, \bibinfo {author} {\bibfnamefont {Q.}~\bibnamefont {Bailey}}, \
  and\ \bibinfo {author} {\bibfnamefont {W.~I.}\ \bibnamefont {Goldburg}},\
  }\bibfield  {title} {\enquote {\bibinfo {title} {Modification of a vortex
  street by a polymer additive},}\ }\href@noop {} {\bibfield  {journal}
  {\bibinfo  {journal} {Phys. Fluids}\ }\textbf {\bibinfo {volume} {13}},\
  \bibinfo {pages} {867--871} (\bibinfo {year} {2001})}\BibitemShut {NoStop}%
\bibitem [{\citenamefont {Zhu}\ and\ \citenamefont {Xi}(2019)}]{zhu2019vortex}%
  \BibitemOpen
  \bibfield  {author} {\bibinfo {author} {\bibfnamefont {L.}~\bibnamefont
  {Zhu}}\ and\ \bibinfo {author} {\bibfnamefont {L.}~\bibnamefont {Xi}},\
  }\bibfield  {title} {\enquote {\bibinfo {title} {Vortex dynamics in low-and
  high-extent polymer drag reduction regimes revealed by vortex tracking and
  conformation analysis},}\ }\href@noop {} {\bibfield  {journal} {\bibinfo
  {journal} {Phys. Fluids}\ }\textbf {\bibinfo {volume} {31}},\ \bibinfo
  {pages} {095103} (\bibinfo {year} {2019})}\BibitemShut {NoStop}%
\bibitem [{\citenamefont {Hopkins}, \citenamefont {Haward},\ and\ \citenamefont
  {Shen}(2020)}]{hopkins2020tristability}%
  \BibitemOpen
  \bibfield  {author} {\bibinfo {author} {\bibfnamefont {C.~C.}\ \bibnamefont
  {Hopkins}}, \bibinfo {author} {\bibfnamefont {S.~J.}\ \bibnamefont {Haward}},
  \ and\ \bibinfo {author} {\bibfnamefont {A.~Q.}\ \bibnamefont {Shen}},\
  }\bibfield  {title} {\enquote {\bibinfo {title} {Tristability in viscoelastic
  flow past side-by-side microcylinders},}\ }\href@noop {} {\bibfield
  {journal} {\bibinfo  {journal} {arXiv preprint arXiv:2010.14749}\ } (\bibinfo
  {year} {2020})}\BibitemShut {NoStop}%
\bibitem [{\citenamefont {Mohammadigoushki}\ and\ \citenamefont
  {Muller}(2016)}]{mohammadigoushki2016sedimentation}%
  \BibitemOpen
  \bibfield  {author} {\bibinfo {author} {\bibfnamefont {H.}~\bibnamefont
  {Mohammadigoushki}}\ and\ \bibinfo {author} {\bibfnamefont {S.~J.}\
  \bibnamefont {Muller}},\ }\bibfield  {title} {\enquote {\bibinfo {title}
  {Sedimentation of a sphere in wormlike micellar fluids},}\ }\href@noop {}
  {\bibfield  {journal} {\bibinfo  {journal} {J. Rheol.}\ }\textbf {\bibinfo
  {volume} {60}},\ \bibinfo {pages} {587--601} (\bibinfo {year}
  {2016})}\BibitemShut {NoStop}%
\bibitem [{\citenamefont {Chen}\ and\ \citenamefont
  {Rothstein}(2004)}]{chen2004flow}%
  \BibitemOpen
  \bibfield  {author} {\bibinfo {author} {\bibfnamefont {S.}~\bibnamefont
  {Chen}}\ and\ \bibinfo {author} {\bibfnamefont {J.~P.}\ \bibnamefont
  {Rothstein}},\ }\bibfield  {title} {\enquote {\bibinfo {title} {Flow of a
  wormlike micelle solution past a falling sphere},}\ }\href@noop {} {\bibfield
   {journal} {\bibinfo  {journal} {J. Non-Newt. Fluid Mech.}\ }\textbf
  {\bibinfo {volume} {116}},\ \bibinfo {pages} {205--234} (\bibinfo {year}
  {2004})}\BibitemShut {NoStop}%
\bibitem [{\citenamefont {Sasmal}(2021)}]{sasmalJFM}%
  \BibitemOpen
  \bibfield  {author} {\bibinfo {author} {\bibfnamefont {C.}~\bibnamefont
  {Sasmal}},\ }\bibfield  {title} {\enquote {\bibinfo {title} {Unsteady motion
  past a sphere translating steadily in wormlike micellar solutions:{A}
  numerical analysis},}\ }\href@noop {} {\bibfield  {journal} {\bibinfo
  {journal} {J. Fluid Mech.}\ }\textbf {\bibinfo {volume} {In press}} (\bibinfo
  {year} {2021})}\BibitemShut {NoStop}%
\bibitem [{\citenamefont {Vasquez}, \citenamefont {McKinley},\ and\
  \citenamefont {Cook}(2007)}]{vasquez2007network}%
  \BibitemOpen
  \bibfield  {author} {\bibinfo {author} {\bibfnamefont {P.~A.}\ \bibnamefont
  {Vasquez}}, \bibinfo {author} {\bibfnamefont {G.~H.}\ \bibnamefont
  {McKinley}}, \ and\ \bibinfo {author} {\bibfnamefont {P.~L.}\ \bibnamefont
  {Cook}},\ }\bibfield  {title} {\enquote {\bibinfo {title} {A network scission
  model for wormlike micellar solutions: {I}. {M}odel formulation and
  viscometric flow predictions},}\ }\href@noop {} {\bibfield  {journal}
  {\bibinfo  {journal} {J. Non-Newt. Fluid Mech.}\ }\textbf {\bibinfo {volume}
  {144}},\ \bibinfo {pages} {122--139} (\bibinfo {year} {2007})}\BibitemShut
  {NoStop}%
\bibitem [{\citenamefont {Cates}(1987)}]{cates1987reptation}%
  \BibitemOpen
  \bibfield  {author} {\bibinfo {author} {\bibfnamefont {M.~E.}\ \bibnamefont
  {Cates}},\ }\bibfield  {title} {\enquote {\bibinfo {title} {Reptation of
  living polymers: dynamics of entangled polymers in the presence of reversible
  chain-scission reactions},}\ }\href@noop {} {\bibfield  {journal} {\bibinfo
  {journal} {Macromolecules}\ }\textbf {\bibinfo {volume} {20}},\ \bibinfo
  {pages} {2289--2296} (\bibinfo {year} {1987})}\BibitemShut {NoStop}%
\bibitem [{\citenamefont {Pipe}\ \emph {et~al.}(2010)\citenamefont {Pipe},
  \citenamefont {Kim}, \citenamefont {Vasquez}, \citenamefont {Cook},\ and\
  \citenamefont {McKinley}}]{pipe2010wormlike}%
  \BibitemOpen
  \bibfield  {author} {\bibinfo {author} {\bibfnamefont {C.~J.}\ \bibnamefont
  {Pipe}}, \bibinfo {author} {\bibfnamefont {N.~J.}\ \bibnamefont {Kim}},
  \bibinfo {author} {\bibfnamefont {P.~A.}\ \bibnamefont {Vasquez}}, \bibinfo
  {author} {\bibfnamefont {L.~P.}\ \bibnamefont {Cook}}, \ and\ \bibinfo
  {author} {\bibfnamefont {G.~H.}\ \bibnamefont {McKinley}},\ }\bibfield
  {title} {\enquote {\bibinfo {title} {Wormlike micellar solutions: {II}.
  {C}omparison between experimental data and scission model predictions},}\
  }\href@noop {} {\bibfield  {journal} {\bibinfo  {journal} {J. Rheol.}\
  }\textbf {\bibinfo {volume} {54}},\ \bibinfo {pages} {881--913} (\bibinfo
  {year} {2010})}\BibitemShut {NoStop}%
\bibitem [{\citenamefont {Zhou}, \citenamefont {McKinley},\ and\ \citenamefont
  {Cook}(2014)}]{zhou2014wormlike}%
  \BibitemOpen
  \bibfield  {author} {\bibinfo {author} {\bibfnamefont {L.}~\bibnamefont
  {Zhou}}, \bibinfo {author} {\bibfnamefont {G.~H.}\ \bibnamefont {McKinley}},
  \ and\ \bibinfo {author} {\bibfnamefont {L.~P.}\ \bibnamefont {Cook}},\
  }\bibfield  {title} {\enquote {\bibinfo {title} {Wormlike micellar solutions:
  {III}. {VCM} model predictions in steady and transient shearing flows},}\
  }\href@noop {} {\bibfield  {journal} {\bibinfo  {journal} {J. Non-Newt. Fluid
  Mech.}\ }\textbf {\bibinfo {volume} {211}},\ \bibinfo {pages} {70--83}
  (\bibinfo {year} {2014})}\BibitemShut {NoStop}%
\bibitem [{\citenamefont {Kalb}, \citenamefont {Cromer}\ \emph
  {et~al.}(2017)\citenamefont {Kalb}, \citenamefont {Cromer} \emph
  {et~al.}}]{kalb2017role}%
  \BibitemOpen
  \bibfield  {author} {\bibinfo {author} {\bibfnamefont {A.}~\bibnamefont
  {Kalb}}, \bibinfo {author} {\bibfnamefont {M.}~\bibnamefont {Cromer}},  \emph
  {et~al.},\ }\bibfield  {title} {\enquote {\bibinfo {title} {Role of chain
  scission in cross-slot flow of wormlike micellar solutions},}\ }\href@noop {}
  {\bibfield  {journal} {\bibinfo  {journal} {Phys. Rev. Fluids}\ }\textbf
  {\bibinfo {volume} {2}},\ \bibinfo {pages} {071301} (\bibinfo {year}
  {2017})}\BibitemShut {NoStop}%
\bibitem [{\citenamefont {Kalb}, \citenamefont {Villasmil-Urdaneta},\ and\
  \citenamefont {Cromer}(2018)}]{kalb2018elastic}%
  \BibitemOpen
  \bibfield  {author} {\bibinfo {author} {\bibfnamefont {A.}~\bibnamefont
  {Kalb}}, \bibinfo {author} {\bibfnamefont {L.~A.}\ \bibnamefont
  {Villasmil-Urdaneta}}, \ and\ \bibinfo {author} {\bibfnamefont
  {M.}~\bibnamefont {Cromer}},\ }\bibfield  {title} {\enquote {\bibinfo {title}
  {Elastic instability and secondary flow in cross-slot flow of wormlike
  micellar solutions},}\ }\href@noop {} {\bibfield  {journal} {\bibinfo
  {journal} {J. Non-Newt. Fluid Mech.}\ }\textbf {\bibinfo {volume} {262}},\
  \bibinfo {pages} {79--91} (\bibinfo {year} {2018})}\BibitemShut {NoStop}%
\bibitem [{\citenamefont {Sasmal}(2020)}]{sasmal2020flow}%
  \BibitemOpen
  \bibfield  {author} {\bibinfo {author} {\bibfnamefont {C.}~\bibnamefont
  {Sasmal}},\ }\bibfield  {title} {\enquote {\bibinfo {title} {Flow of wormlike
  micellar solutions through a long micropore with step expansion and
  contraction},}\ }\href@noop {} {\bibfield  {journal} {\bibinfo  {journal}
  {Phys. Fluids}\ }\textbf {\bibinfo {volume} {32}},\ \bibinfo {pages} {013103}
  (\bibinfo {year} {2020})}\BibitemShut {NoStop}%
\bibitem [{\citenamefont {Mohammadigoushki}\ \emph {et~al.}(2019)\citenamefont
  {Mohammadigoushki}, \citenamefont {Dalili}, \citenamefont {Zhou},\ and\
  \citenamefont {Cook}}]{mohammadigoushki2019transient}%
  \BibitemOpen
  \bibfield  {author} {\bibinfo {author} {\bibfnamefont {H.}~\bibnamefont
  {Mohammadigoushki}}, \bibinfo {author} {\bibfnamefont {A.}~\bibnamefont
  {Dalili}}, \bibinfo {author} {\bibfnamefont {L.}~\bibnamefont {Zhou}}, \ and\
  \bibinfo {author} {\bibfnamefont {P.}~\bibnamefont {Cook}},\ }\bibfield
  {title} {\enquote {\bibinfo {title} {Transient evolution of flow profiles in
  a shear banding wormlike micellar solution: {E}xperimental results and a
  comparison with the {VCM} model},}\ }\href@noop {} {\bibfield  {journal}
  {\bibinfo  {journal} {Soft Matt.}\ }\textbf {\bibinfo {volume} {15}},\
  \bibinfo {pages} {5483--5494} (\bibinfo {year} {2019})}\BibitemShut {NoStop}%
\bibitem [{\citenamefont {Weller}\ \emph {et~al.}(1998)\citenamefont {Weller},
  \citenamefont {Tabor}, \citenamefont {Jasak},\ and\ \citenamefont
  {Fureby}}]{wellerOpenFOAM}%
  \BibitemOpen
  \bibfield  {author} {\bibinfo {author} {\bibfnamefont {H.~G.}\ \bibnamefont
  {Weller}}, \bibinfo {author} {\bibfnamefont {G.}~\bibnamefont {Tabor}},
  \bibinfo {author} {\bibfnamefont {H.}~\bibnamefont {Jasak}}, \ and\ \bibinfo
  {author} {\bibfnamefont {C.}~\bibnamefont {Fureby}},\ }\bibfield  {title}
  {\enquote {\bibinfo {title} {A tensorial approach to computational continuum
  mechanics using object-oriented techniques},}\ }\href@noop {} {\bibfield
  {journal} {\bibinfo  {journal} {Com. Phys.}\ }\textbf {\bibinfo {volume}
  {12}},\ \bibinfo {pages} {620--631} (\bibinfo {year} {1998})}\BibitemShut
  {NoStop}%
\bibitem [{\citenamefont {Pimenta}\ and\ \citenamefont
  {Alves}(2016)}]{rheoTool}%
  \BibitemOpen
  \bibfield  {author} {\bibinfo {author} {\bibfnamefont {F.}~\bibnamefont
  {Pimenta}}\ and\ \bibinfo {author} {\bibfnamefont {M.}~\bibnamefont
  {Alves}},\ }\href@noop {} {\enquote {\bibinfo {title} {rheotool},}\ }\bibinfo
  {howpublished} {\url{https://github.com/fppimenta/rheoTool}} (\bibinfo {year}
  {2016})\BibitemShut {NoStop}%
\bibitem [{\citenamefont {Ajiz}\ and\ \citenamefont
  {Jennings}(1984)}]{ajiz1984robust}%
  \BibitemOpen
  \bibfield  {author} {\bibinfo {author} {\bibfnamefont {M.~A.}\ \bibnamefont
  {Ajiz}}\ and\ \bibinfo {author} {\bibfnamefont {A.}~\bibnamefont
  {Jennings}},\ }\bibfield  {title} {\enquote {\bibinfo {title} {A robust
  incomplete choleski-conjugate gradient algorithm},}\ }\href@noop {}
  {\bibfield  {journal} {\bibinfo  {journal} {Int. J. Num. Methods Eng.}\
  }\textbf {\bibinfo {volume} {20}},\ \bibinfo {pages} {949--966} (\bibinfo
  {year} {1984})}\BibitemShut {NoStop}%
\bibitem [{\citenamefont {Lee}, \citenamefont {Zhang},\ and\ \citenamefont
  {Lu}(2003)}]{lee2003incomplete}%
  \BibitemOpen
  \bibfield  {author} {\bibinfo {author} {\bibfnamefont {J.}~\bibnamefont
  {Lee}}, \bibinfo {author} {\bibfnamefont {J.}~\bibnamefont {Zhang}}, \ and\
  \bibinfo {author} {\bibfnamefont {C.~C.}\ \bibnamefont {Lu}},\ }\bibfield
  {title} {\enquote {\bibinfo {title} {Incomplete {LU} preconditioning for
  large scale dense complex linear systems from electromagnetic wave scattering
  problems},}\ }\href@noop {} {\bibfield  {journal} {\bibinfo  {journal} {J.
  Comp. Phys.}\ }\textbf {\bibinfo {volume} {185}},\ \bibinfo {pages}
  {158--175} (\bibinfo {year} {2003})}\BibitemShut {NoStop}%
\bibitem [{\citenamefont {Alves}, \citenamefont {Oliveira},\ and\ \citenamefont
  {Pinho}(2003)}]{alves2003convergent}%
  \BibitemOpen
  \bibfield  {author} {\bibinfo {author} {\bibfnamefont {M.~A.}\ \bibnamefont
  {Alves}}, \bibinfo {author} {\bibfnamefont {P.~J.}\ \bibnamefont {Oliveira}},
  \ and\ \bibinfo {author} {\bibfnamefont {F.~T.}\ \bibnamefont {Pinho}},\
  }\bibfield  {title} {\enquote {\bibinfo {title} {A convergent and universally
  bounded interpolation scheme for the treatment of advection},}\ }\href@noop
  {} {\bibfield  {journal} {\bibinfo  {journal} {Int. J. Num. Methods Fluids}\
  }\textbf {\bibinfo {volume} {41}},\ \bibinfo {pages} {47--75} (\bibinfo
  {year} {2003})}\BibitemShut {NoStop}%
\bibitem [{\citenamefont {Haward}, \citenamefont {Hopkins},\ and\ \citenamefont
  {Shen}(2020)}]{haward2020asymmetric}%
  \BibitemOpen
  \bibfield  {author} {\bibinfo {author} {\bibfnamefont {S.~J.}\ \bibnamefont
  {Haward}}, \bibinfo {author} {\bibfnamefont {C.~C.}\ \bibnamefont {Hopkins}},
  \ and\ \bibinfo {author} {\bibfnamefont {A.~Q.}\ \bibnamefont {Shen}},\
  }\bibfield  {title} {\enquote {\bibinfo {title} {Asymmetric flow of polymer
  solutions around microfluidic cylinders: {I}nteraction between shear-thinning
  and viscoelasticity},}\ }\href@noop {} {\bibfield  {journal} {\bibinfo
  {journal} {J. Non-Newt. Fluid Mech.}\ }\textbf {\bibinfo {volume} {278}},\
  \bibinfo {pages} {104250} (\bibinfo {year} {2020})}\BibitemShut {NoStop}%
\bibitem [{\citenamefont {Pakdel}\ and\ \citenamefont
  {McKinley}(1996)}]{pakdel1996elastic}%
  \BibitemOpen
  \bibfield  {author} {\bibinfo {author} {\bibfnamefont {P.}~\bibnamefont
  {Pakdel}}\ and\ \bibinfo {author} {\bibfnamefont {G.~H.}\ \bibnamefont
  {McKinley}},\ }\bibfield  {title} {\enquote {\bibinfo {title} {Elastic
  instability and curved streamlines},}\ }\href@noop {} {\bibfield  {journal}
  {\bibinfo  {journal} {Phys. Rev. Lett.}\ }\textbf {\bibinfo {volume} {77}},\
  \bibinfo {pages} {2459} (\bibinfo {year} {1996})}\BibitemShut {NoStop}%
\bibitem [{\citenamefont {McKinley}, \citenamefont {Pakdel},\ and\
  \citenamefont {{\"O}ztekin}(1996)}]{mckinley1996rheological}%
  \BibitemOpen
  \bibfield  {author} {\bibinfo {author} {\bibfnamefont {G.~H.}\ \bibnamefont
  {McKinley}}, \bibinfo {author} {\bibfnamefont {P.}~\bibnamefont {Pakdel}}, \
  and\ \bibinfo {author} {\bibfnamefont {A.}~\bibnamefont {{\"O}ztekin}},\
  }\bibfield  {title} {\enquote {\bibinfo {title} {Rheological and geometric
  scaling of purely elastic flow instabilities},}\ }\href@noop {} {\bibfield
  {journal} {\bibinfo  {journal} {J. Non-Newt. Fluid Mech.}\ }\textbf {\bibinfo
  {volume} {67}},\ \bibinfo {pages} {19--47} (\bibinfo {year}
  {1996})}\BibitemShut {NoStop}%
\bibitem [{\citenamefont {Fardin}\ and\ \citenamefont
  {Lerouge}(2012)}]{fardin2012instabilities}%
  \BibitemOpen
  \bibfield  {author} {\bibinfo {author} {\bibfnamefont {M.-A.}\ \bibnamefont
  {Fardin}}\ and\ \bibinfo {author} {\bibfnamefont {S.}~\bibnamefont
  {Lerouge}},\ }\bibfield  {title} {\enquote {\bibinfo {title} {Instabilities
  in wormlike micelle systems},}\ }\href@noop {} {\bibfield  {journal}
  {\bibinfo  {journal} {The European Phys. J. E}\ }\textbf {\bibinfo {volume}
  {35}},\ \bibinfo {pages} {1--29} (\bibinfo {year} {2012})}\BibitemShut
  {NoStop}%
\end{thebibliography}%

\end{document}